\pdfoutput=1
\documentclass[floatfix,nofootinbib,twocolumn,english,pra,aps]{revtex4}
\usepackage{silence}
\usepackage[cal=boondox]{mathalfa}
\usepackage{mathtools}
\usepackage{graphicx}
\usepackage{babel}
\makeatletter
\pdfpageheight\paperheight
\pdfpagewidth\paperwidth

\begin{document}
\title{\noindent Entropy bounds from quantum thermodynamics}
\author{\noindent Massimo Giovannini}
\email{massimo.giovannini@cern.ch}
\affiliation{\noindent Department of Physics, CERN, 1211 Geneva 23, Switzerland and INFN, Section of Milan-Bicocca, 20126 Milan, Italy}
\begin{abstract}
\noindent 
Within an inherently classical perspective, there is always an unavoidable energy cost associated with the information deletion and this common lore is at the heart of the Landauer's conjecture that does not impose, per se, any relevant limit on the information acquisition. Although such a mindset should generally apply to systems of any size, its quantum mechanical implications are particularly intriguing and, for this reason, we examine here a minimal physical structure where the system and the environment are described, respectively, by a pair of quantum oscillators coupled by an appropriate Hermitian interaction able to amplify the entropy of the initial state. Since at the onset of the dynamical evolution the system is originally in a pure state, its entropy variation is always positive semidefinite and the Landauer's conjecture should not impose any constraint. Nonetheless, provided the quantum amplification is effective, it turns out that the entropy variation of the system always undershoots the heat transferred to the environment. When the initial thermal state of the environment is characterized by a chemical potential, the entropy growth is bounded both by the particles and by the heat flowing to the environment. The limits deduced in the quantum thermodynamical framework are also scrutinized from a field theory standpoint where species of different spins are copiously produced (especially in a cosmological context) thanks to the rapid variation of the space-time curvature.
\end{abstract}
\maketitle

\renewcommand{\theequation}{1.\arabic{equation}}
\setcounter{equation}{0}
\section{Introduction}
\label{sec1}
According to the current wisdom, a fundamental energy (or heat) cost 
must always be associated with the information erasure and this perspective supports the view that information is unavoidably linked to an empirical representation \cite{ONE}.
There also exist complementary viewpoints suggesting 
that mathematical and geometrical structures should not depend upon the specific devices employed 
for their computational assessment \cite{TWO}. Between these two interesting perspectives the former 
is more common than the latter when applying thermodynamics to information theory, or vice versa.
In particular the so-called Landauer conjecture \cite{THREE,FOUR,FIVE,SIX,SEVEN} would suggest that the irreversible deletion of one bit of information demands an energy 
cost larger than (or equal to) $\kappa_{B} \,T_{e} \ln{2}$ where $\kappa_{B}$ is the Boltzmann constant and $T_{e}$ is the temperature of the environment.  This formulation implicitly propounds that there should not be any restriction connected to the acquisition of information; in other words the energy cost associated with the information erasure is unavoidable but the information acquisition is not constrained, at least 
within the purely classical perspective where, among other things, the Landauer's conjecture provides a possible set of solutions to the Maxwell's paradox \cite{THREE,FOUR} (see also \cite{EIGHT} for a review). In more quantitative terms, if $\Delta Q_{e} \geq 0$ is the amount of heat flowing to the environment and $\Delta S_{s}$ is the variation of the entropy of the system 
the Landauer's conjecture stipulates that 
\begin{equation}
\Delta Q_{e} \geq - \kappa_{B} \, T_{e}\, \Delta S_{s}.
\label{INT1}
\end{equation}
 When information is erased from the system this means that $\Delta S_{s} <0$ and Eq. (\ref{INT1}) demands a specific energy cost; for instance, to erase one bit of information $\Delta S_{s} = - \ln{2}$ 
and Eq. (\ref{INT1}) would imply that the heat flowing to the environment 
should be, at least, comparable with $\kappa_{B} \,T_{e} \,\ln{2}$.

The bound of Eq. (\ref{INT1}) holds in spite of the size of the underlying 
physical structures and it should then apply, in particular, to {\em any} quantum system. Furthermore Eq. (\ref{INT1}) 
is effectively constraining only when $\Delta S_{s} <0$ but it
should not imply any limit when $\Delta S_{s} >0$: in this case 
Eq. (\ref{INT1}) is obviously verified since a positive semi-definite 
increment must always exceed a negative contribution. In this paper, following a 
recent suggestion \cite{NINE}, we feel that it is CORRECTEDboth necessary and interesting to discuss at length 
one of the simplest  quantum mechanical playgrounds where the potential bounds on the entropy growth could be quantitatively scrutinized. The system is initially in a pure state given by a quantum oscillator in the vacuum while the environment is instead constituted by a further oscillator in a mixed state whose density operator is characterized by Bose-Einstein weights. The Hermitian interactions between the system and the environment follow
from the requirement that the von Neumann entropy of the initial state is amplified and this may occur, as we shall demonstrate, in the quantum mechanical description of parametric amplification \cite{TEN,ELEVEN,TWELVE} (see also \cite{THIRTEEN,FOURTEEN}). After revisiting the standard tenets of parametric amplification within a quantum thermodynamical perspective, it will be shown that the variation of the  entropy of the 
system is always positive (i.e. $\Delta S_{s} \geq 0$) but it is 
nonetheless constrained by the heat transfer according to the following bound
\begin{equation}
\kappa_{B} \, T_{e} \,\Delta S_{s} \leq \Delta Q_{e}.
\label{INT2}
\end{equation}
From the quantum mechanical viewpoint both Eqs. (\ref{INT1}) and (\ref{INT2}) 
hold under the assumption that the environment is initially in a thermal state with temperature 
$T_{e}$ while the system is in a pure state (e.g. the vacuum). 
We shall additionally argue that  Eq. (\ref{INT2}) can be generalized 
to the field theoretical situation where particles are produced because of the 
variation of the space-time curvature \cite{FIFTEEN,SIXTEEN} in the cases 
of curvature phonons \cite{SEVENTEEN,EIGHTEEN} and of gravitons \cite{NINETEEN,TWENTY,TWENTYONE,TWENTYONEa} (see also \cite{TWENTYTWO,TWENTYTWOa,TWENTYTWOb,TWENTYTWOc,TWENTYTWOd}).

Before plunging into the details of the discussion we stress that, although in Eqs. (\ref{INT1})--(\ref{INT2}) the Boltzmann constant $\kappa_{B}$ has been included, we are going to adopt hereunder the natural system of units where $\hbar= c= \kappa_{B} =1$. In this system the Newton's constant $G$, the Planck length $\ell_{P}$ and the Planck mass $M_{P}$ are then related as $\ell_{P} = \sqrt{8\pi}/M_{P}= 8\pi G$. In the remaining three subsections of this introduction we shall present, step by step, the classical viewpoint leading to Eq. (\ref{INT1}), the basic tenets of the quantum thermodynamical approach and the overall logic of this investigation.
 
\subsection{The classical perspective}
In classical information theory the Shannon entropy measures the lack of specific knowledge one 
has about a given system \cite{TWENTYTHREE,TWENTYFOUR,TWENTYFIVE}. If we suppose the system is 
given by a particle that can be in two different energy levels (e.g. $+$ and $-$)
with equal probability (i.e. $p^{(in)}_{+}= p^{(in)}_{-} =1/2$) the Shannon entropy associated with this initial situation is  
\begin{equation}
S_{s}^{(in)} = - p^{(in)}_{+} \ln{p^{(in)}_{+}} - p^{(in)}_{-} \, \ln{p^{(in)}_{-}} = \ln{2}.
\label{INT3}
\end{equation}
Let us now imagine that the final state of the system always excludes 
the presence of the particle in the lower state so that $S_{s}^{(fin)} =0$
since $p^{(fin)}_{+} =1$ and $p^{(fin)}_{-} =0$. In this process the information 
of the initial state has been erased since any potential knowledge about the initial 
state of the particle is forever obliterated. Moreover, as expected, the variation of the Shannon entropy is  
negative 
\begin{equation}
\Delta S_{s} = S_{s}^{(fin)} - S_{s}^{(in)} = - \ln{2} <0,
\label{INT4}
\end{equation}
and the Landauer's conjecture of Eq. (\ref{INT1}) would then demand that 
for the deletion of this one bit of information the minimal 
energy cost is given by\footnote{To simplify the notation and to avoid the proliferation of indices 
we shall always denote by $T$ the temperature of the environment (i.e. $T_{e} = T$); since there is no other 
temperature appearing in the discussion this notation is unambiguous.} $\Delta Q_{e} \geq T \, \ln{2}$.  
Generally speaking, if a certain state $p_{n}$ has probability $1$ this means 
that the {\em surprise} tends to zero where the surprise, in classical information theory, is simply the 
natural logarithm of $1/p_{n}$ \cite{TWENTYFIVE}, i.e. $- \ln{p_{n}}$. This also means that after erasing 
the information, the surprise and the entropy decrease since the Shannon entropy is, in practice, the average surprise of a given state.
Conversely, more surprise implies instead an increase of the information and of the entropy.
Since the removal of information requires a decrease of the entropy of the system, when $\Delta S_{s} >0$ 
the condition imposed by Eq. (\ref{INT1}) does not seem restrictive: a physical quantity 
which is positive semi-definite (i.e. $\Delta Q_{e}\geq 0$) always exceeds a negative 
contribution (i.e. $- T \Delta S_{e}$). Once more the 
deletion of information costs energy whereas its acquisition remains unconstrained, at least within 
the classical perspective.  Even if the potential saturation of the bound given by Eq. (\ref{INT1}) is under debate \cite{TWENTYSIX}, the classical logic behind the Landauer's conjecture has been experimentally verified in a number of different frameworks \cite{TWENTYSEVEN,TWENTYEIGHT} and will not be questioned hereunder.

\subsection{The quantum mechanical perspective}
Since Eq. (\ref{INT1}) applies in spite of the physical dimensions of the underlying structures,  
 it should be particularly relevant for any quantum system \cite{EIGHT} where  the Landauer's conjecture is customarily formulated by introducing the Hilbert spaces of the system and of the environment \cite{TWENTYNINE}.
Both the {\em system} and the {\em environment} are described by the corresponding 
Hamiltonian operators that we shall denote, respectively, by $\widehat{H}_{s}$ and 
$\widehat{H}_{e}$. The interaction Hamiltonian (denoted by $\widehat{H}_{s\,e}$) 
is usually expected to be Hermitian so that the final and the initial density matrices are in fact 
related by the unitary operator $\widehat{{\mathcal U}}( t_{in}, \, t_{fin})$ accounting for the global 
time evolution of the system and of the environment:
\begin{equation}
\widehat{\rho}(t_{fin}) = \widehat{{\mathcal U}}(t_{in},\,t_{fin}) \, \widehat{\rho}(t_{in}) \, \widehat{{\mathcal U}}^{\dagger}( t_{in},\,t_{fin}),
\label{INT5}
\end{equation}
where $\widehat{\rho}(t)$ denotes the total density matrix encompassing the system and the environment. Occasionally $\widehat{\rho}(t)$ is called proper density operator \cite{THIRTY} as opposed to the 
improper one (obtained by tracing the total density matrix either over the degrees of freedom 
of the system or over the ones of the environment). Bearing in mind this standard terminology, the quantum mechanical formulation of the Landauer's conjecture stipulates that the 
system and the environment are {\it (i)} {\em initially} uncorrelated and that 
{\it (ii)} the environment is in a thermal state \cite{TWENTYNINE}.  The absence 
of correlations at $t_{in}$ demands that $\widehat{\rho}(t_{in}) = \widehat{\rho}_{s}(t_{in})\otimes\widehat{\rho}_{e}(t_{in})$; thus Eq. (\ref{INT5}) can also be written as
\begin{equation}
\widehat{\rho}(t_{fin}) = \widehat{{\mathcal U}}(t_{in},\,t_{fin}) \,\widehat{\rho}_{s}(t_{in})\otimes\widehat{\rho}_{e}(t_{in}) \, \widehat{{\mathcal U}}^{\dagger}(t_{in},\,t_{fin}).
\label{INT6}
\end{equation}
The lack of initial correlations between the system and the environment does not imply 
that the two shall remain uncorrelated later on. On the contrary we are specifically 
interested in the situation where the entropy of the system increases between the initial and the 
final stages of the dynamical evolution. At any generic time  
$t$ the reduced (or improper) density matrices are obtained by tracing $\widehat{\rho}(t)$
 with respect to the degrees of freedom either of the system or of the 
environment. This means, in particular, that at $t_{fin}$
\begin{equation}
\widehat{\rho}_{s}(t_{fin}) = \mathrm{Tr}_{e} \bigl[ \widehat{\rho}(t_{fin})\bigr],\quad
\widehat{\rho}_{e}(t_{fin}) = \mathrm{Tr}_{s} \bigl[ \widehat{\rho}(t_{fin})\bigr].
\label{INT7}
\end{equation}
where $\mathrm{Tr}_{e}$ and $\mathrm{Tr}_{s}$ indicate, respectively, the traces 
over the degrees of freedom of the environment and of the system.

The corresponding von Neumann entropies and the associated heat transfer follow
from the reduced density matrices of Eq. (\ref{INT7}); for instance the von Neumann entropy of the system is given by:
\begin{equation}
S_{s}(t_{fin}) = - \mathrm{Tr}[\widehat{\rho}_{s}(t_{fin}) \, \ln{\widehat{\rho}_{s}(t_{fin})}],
\label{INT8}
\end{equation}
where the trace is now performed directly over the dynamical degrees of freedom 
of the system. Similarly the expectation value of the heat transferred to the environment 
will be
\begin{equation}
Q_{e}(t_{fin}) = \mathrm{Tr}[\widehat{H}_{e}(t_{fin}) \widehat{\rho}_{e}(t_{fin})].
\label{INT9}
\end{equation}
In terms of Eqs. (\ref{INT8})--(\ref{INT9}) 
the variation of the von Neumann entropy of the system and the heat flowing to the 
environment are given, respectively, by 
\begin{eqnarray}
\Delta S_{s} &=& S_{s}\bigl[\widehat{\rho}_{s}(t_{fin})\bigr] - S_{s}\bigl[\widehat{\rho}_{s}(t_{in})\bigr],
\label{INT10}\\
\Delta Q_{e} &=& Q_{e}\bigl[\widehat{\rho}_{e}(t_{fin})\bigr] - Q_{e}\bigl[\widehat{\rho}_{e}(t_{in})\bigr].
\label{INT11}
\end{eqnarray}
The heat transferred to the environment following from Eq. (\ref{INT11}) 
must coincide with the expectation values of the Hamiltonian of the environment 
(evaluated in the Schr\"odinger description) between 
the final and the initial states of the evolution:
\begin{equation}
\Delta Q_{e} = \langle fin| \widehat{H}_{e}(t_{in}) |fin \rangle - \langle in| \widehat{H}_{e}(t_{in}) |in \rangle,
\label{INT11a}
\end{equation}
where now the Hamiltonian is evaluated at $t_{in}$.
It can finally happen, in the present context, that the initial state allows for the presence of a chemical potential.
In this situation we may expect that also the particles flow to the environment and the corresponding 
increment, in analogy with Eq. (\ref{INT11a}), is 
\begin{equation}
\Delta N_{e} = \langle fin|  \widehat{N}_{e}(t_{in})  |fin \rangle - \langle in| \widehat{N}_{e}(t_{in}) |in \rangle,
\label{INT13}
\end{equation}
where $\widehat{N}_{e}$ is now the averaged multiplicity of the environment. The inclusion 
of $\Delta N_{e}$ is not usually considered in the quantum formulation of the Landauer's conjecture 
but it might be relevant in our present discussion, as we are going to argue when discussing the nature 
of the entropy bounds.

\subsection{Quantum parametric amplification}
From the quantum mechanical viewpoint the Landauer's conjecture becomes then a restriction 
on the mutual relation between the energy increment of the environment and the variation of the 
von Neumann entropy of the system.
The explicit form of this restriction is exactly given by Eq. (\ref{INT1}) where however 
$\Delta Q_{e}$ and $\Delta S_{s}$ are now computed from the expectation values 
of the Hamiltonian and from the von Neumann entropy given, respectively, by Eqs. (\ref{INT10})--(\ref{INT11}).

We point out in this investigation that quantum mechanics does not necessarily dictate that the growth of the 
entropy of the system (i.e. $\Delta S_{s} >0$) must remain completely unconstrained.  
On the contrary the purpose of this analysis is to show that the variation of the entropy of the system 
may well be bounded by the heat transfer as suggested by Eq. (\ref{INT2}).
For the actual derivation of the bound expressed by Eq. (\ref{INT2}) we shall be examining here the minimal 
situation where the system and the environment correspond, respectively, to a pair of quantum oscillators with frequencies $\omega_{s}$ 
and $\omega_{e}$
\begin{eqnarray}
&& \widehat{H}_{s} = \omega_{s} (\widehat{s}^{\dagger} \, \widehat{s} + 1/2) = \omega_{s} \bigl(\widehat{N}_{s} +1/2\bigr),
\label{INT14}\\ 
&&\widehat{H}_{e} = \omega_{e} (\widehat{e}^{\dagger} \, \widehat{e} + 1/2)= \omega_{e} \bigl(\widehat{N}_{e} +1/2\bigr),
\label{INT15}
\end{eqnarray}
where $[\,\widehat{s},\,\widehat{e}\,]=0$ and similarly for the associated number operators 
$\widehat{N}_{s} $ and $\widehat{N}_{e}$, i.e.  $[\,\widehat{N}_{s},\,\widehat{N}_{e}\,]=0$.
The presence of the zero-point energy in the Hamiltonians is not essential but we shall anyway stick to 
the expressions of Eqs. (\ref{INT14})--(\ref{INT15}). 

To avoid spurious effects that might influence the conclusions,
the interplay between the system and the environment should satisfy the following three plausible
requirements: {\it (i)} the mutual interactions must be Hermitian to avoid violations of the unitary evolution; {\it (ii)} they must  
lead to a positive heat transfer to the environment (i.e. $\Delta Q_{e} \geq 0$) 
and {\it (iii)} they must also yield a positive entropy variation of the system (i.e. $\Delta S_{s} \geq 0$).
In case these three conditions are verified it remains to be understood if $\Delta S_{s}$
 is limited (as suggested by Eq. (\ref{INT2})) or if it is unbounded, 
as implied by the classical situation and by the conventional form of the Landauer's conjecture. 
As we shall see, the conditions spelled out above are potentially verified in the context of the quantum theory 
of parametric amplification \cite{TEN,ELEVEN,TWELVE,THIRTEEN,FOURTEEN}. However, although the creation and the annihilation operators of Eqs. (\ref{INT14})--(\ref{INT15}) may be connected to the signal and idler modes, the quantum thermodynamical analysis pursued here suggests a different interpretation that is dictated by the correlation properties of the initial  density operators.

After these general premises the layout of this investigation is, in short, the following. In section \ref{sec2} 
the coupled quantum evolutions of system and of the  environment are discussed in the Heisenberg description when the free Hamiltonians are represented by Eqs. (\ref{INT14})--(\ref{INT15}) and with particular attention to the symmetries of the problem. In section \ref{sec3} we focus on the specific form 
of the reduced density operators and connect their properties with  the irreducible representations of the $SU(1,1)$ group. The heat flow and the entropy are directly computed and analyzed in section \ref{sec4} while section \ref{sec5} is devoted to the scrutiny of the physical bounds connecting the variation of the entropy with the heat flow. At the end of section \ref{sec5} we also comment on the possibile presence of a chemical potential.
In section \ref{sec6}, motivated by the cosmological evolution,  we examine the 
quantum mechanical considerations leading to the bound of Eq. (\ref{INT2}) from the standpoint of field theory, with 
particular attention to the production of spin $0$ and spin $2$ particles in curved background geometries. 
Section \ref{sec7} contains our concluding remarks. 

\renewcommand{\theequation}{2.\arabic{equation}}
\setcounter{equation}{0}
\section{The system and the environment}
\label{sec2} 
\subsection{The interaction Hamiltonian}
The requirements formulated at the end of  Section \ref{sec1} imply that  the interaction Hamiltonian must be Hermitian to avoid 
non-unitary effects coming from the coupled evolution of the system and of the environment; moreover the 
explicit quantum mechanical forms of $\Delta S_{s}$ and $\Delta Q_{e}$ (see Eqs. (\ref{INT10})--(\ref{INT11}) and (\ref{INT11a})) is positive 
semi-definite provided $\widehat{H}_{s\,e}$ does not commute with the sum of the number operators of the system and of the environment. A  general form of $\widehat{H}_{s\,e}$ satisfying these three independent physical conditions turns out to be 
\cite{TEN,ELEVEN,TWELVE} (see also \cite{THIRTEEN,FOURTEEN}):
\begin{equation}
\widehat{H}_{s\,e} = g(t) \,\widehat{s}^{\dagger} \, \widehat{e}^{\dagger} \,\, e^{ - i \omega \, t} + g^{\ast}(t) \,\widehat{e} \, \widehat{s} \,\,e^{ i \omega \, t},
\label{Q3}
\end{equation}
where $g(t)= q(t)\, e^{i\, \theta_{in}}$ and $\omega = \omega_{s} + \omega_{e}$.
 The evolution of $\widehat{s}$ and $\widehat{e}$ follows from the total Hamiltonian obtained from the sum of Eqs. (\ref{INT14})--(\ref{INT15}) and (\ref{Q3}), i.e. 
\begin{equation}
\widehat{H}= \widehat{H}_{s}  + \widehat{H}_{e} + \widehat{H}_{s\,e}. 
\label{Q4}
\end{equation}
In the Heisenberg description the evolution of the operators is given by:
\begin{eqnarray}
\frac{d \widehat{s}}{d t} &=& i\, [\widehat{H}, \widehat{s}] = - i \, \omega_{s} \, \widehat{s} - i \, g\, e^{- i\, \omega \, t} \,\, \widehat{e}^{\,\dagger},
\label{Q5}\\
\frac{d \widehat{e}}{d t} &=& i\, [\widehat{H}, \widehat{e}] = - i \, \omega_{e} \, \widehat{e} - i \,g\, e^{- i\, \omega \, t} \,\, \widehat{s}^{\,\dagger}.
\label{Q6}
\end{eqnarray}
The solution of Eqs. (\ref{Q5})--(\ref{Q6}) can be expressed in terms of four 
complex functions, namely:
\begin{eqnarray}
\widehat{s}(t) &=& u_{s}(t) \, \widehat{{\mathcal s}} - v_{s}(t) \, \widehat{{\mathcal e}}^{\,\dagger}, 
\label{Q70}\\
\widehat{e}(t) &=& u_{e}(t) \, \widehat{{\mathcal e}} - v_{e}(t) \, \widehat{{\mathcal s}}^{\,\dagger}, 
\label{Q7}
\end{eqnarray}
where, to avoid possible confusions, the following notations have been adopted 
\begin{equation}
\widehat{{\mathcal s}} = \widehat{s}(t_{in}), \, \qquad \widehat{{\mathcal e}} = \widehat{e}(t_{in}),
\label{Q7a}
\end{equation}
and will be enforced throughout the whole discussion; the calligraphic operators $\widehat{{\mathcal s}}$ and $\widehat{{\mathcal e}}$ 
actually appear in the Schr\"odinger description that plays a relevant r\^ole in the determination of the asymptotic states (see, in this respect, section \ref{sec3}). We should also stress that, because of the unitarity of the evolution,  the functions $[u_{e}(t),\,v_{e}(t)]$ and $[u_{s}(t),\,v_{s}(t)]$ must satisfy, at any stage of the dynamical evolution, the conditions $| u_{s}(t)|^2 - |v_{s}(t)|^2 =1$ and $| u_{e}(t)|^2 - |v_{e}(t)|^2 =1$. 

\subsection{Solutions of the coupled evolution}
After inserting Eqs. (\ref{Q70})--(\ref{Q7}) into Eqs. (\ref{Q5})--(\ref{Q6})  
the evolution of the four unknown functions becomes\footnote{In principle the $4$ complex functions 
subjected to the conditions $| u_{s}(t)|^2 - |v_{s}(t)|^2 =1$ and $| u_{e}(t)|^2 - |v_{e}(t)|^2 =1$ should 
be equivalent to $6$ real functions. However as a result of the dynamical evolution the system (\ref{Q8})--(\ref{Q9})
depends on one real function and $5$ real phases that can be reduced to $3$ by fixing two of them.}
\begin{eqnarray}
&& \frac{d u_{s}}{d t} = - i\, \omega_{s} \, u_{s} + i \, g(t) e^{-i \, \omega\, t} \, v_{e}^{\ast},
\nonumber\\
&& \frac{d v_{s}}{d t} = - i\, \omega_{s} \, v_{s} + i \, g(t) e^{-i \, \omega\, t} \, u_{e}^{\ast},
\label{Q8}\\
&& \frac{d u_{e}}{d t} = - i\, \omega_{e} \, u_{e} + i \, g(t) e^{-i \, \omega\, t} \, v_{s}^{\ast},
\nonumber\\
&& \frac{d v_{e}}{d t} = - i\, \omega_{e} \, v_{e} + i \, g(t) e^{-i \, \omega\, t} \, u_{s}^{\ast},
\label{Q9}
\end{eqnarray}
and the solutions of Eqs. (\ref{Q8})--(\ref{Q9}) are:
\begin{eqnarray}
u_{s}(t) = e^{i(\vartheta_{s}-\omega_{s}\,t)}\cosh{r},\,\,v_{s}(t)= e^{i(\overline{\vartheta}_{s}- \omega_{s} \, t)}\sinh{r},
\nonumber\\
u_{e}(t) =e^{i(\vartheta_{e}- \omega_{e}\,t)}\cosh{r},\,\,v_{e}(t) = e^{i(\overline{\vartheta}_{e} - \omega_{e} \, t)}\sinh{r},
\nonumber
\end{eqnarray}
where $d r/dt = q(t)$ while 
\begin{eqnarray}
 \overline{\vartheta}_{e} = \theta - \vartheta_{s},\qquad \overline{\vartheta}_{s} = \theta - \vartheta_{e}.
\label{Q12}
\end{eqnarray}
In Eq. (\ref{Q12}) we introduced the new variable $\theta= (\theta_{in} +\pi/2)$ which is going to control 
the phases of the final solution; indeed, without loss of generality, we can always set $\vartheta_{s}= \vartheta_{e} =0$ 
so that the evolution of the operators $\widehat{s}(t)$ and 
$\widehat{e}(t)$ can be ultimately expressed as:
\begin{eqnarray}
\widehat{s}(t) &=& e^{- i\, \omega_{s} t}\bigl[ \cosh{r} \, \widehat{{\mathcal s}} - \sinh{r} \, e^{i \theta} \, \widehat{{\mathcal e}}^{\dagger}\bigr],
\label{Q13}\\
\widehat{e}(t) &=& e^{- i\, \omega_{e} t}\bigl[ \cosh{r} \, \widehat{{\mathcal e}} - \sinh{r} \, e^{i \theta} \, \widehat{{\mathcal s}}^{\dagger}\bigr].
\label{Q14}
\end{eqnarray}
When $\vartheta_{s}\neq 0$ and $\vartheta_{e} \neq 0$ the further contributions can always be absorbed 
in a redefinition of the overall phases appearing outside the square brackets in Eqs. (\ref{Q13})--(\ref{Q14}).
It is now convenient to introduce a pair of unitary operators \cite{THIRTYONE,THIRTYTWO,THIRTYTHREE} denoted hereunder  by
$\widehat{{\mathcal R}}(\delta)$ and $\widehat{\Sigma}(z)$:
\begin{eqnarray}
&& \widehat{{\mathcal R}}(\delta) = e^{- i\, \delta_{s} \,\widehat{{\mathcal s}}^{\dagger} \widehat{{\mathcal s}}
- i \,\delta_{e} \,\widehat{{\mathcal e}}^{\dagger} \widehat{{\mathcal e}}}
\label{Q15}\\
&& \widehat{\Sigma}(z)=e^{z^{\ast}\,\widehat{{\mathcal s}} \,\widehat{{\mathcal e}} - z \, \widehat{{\mathcal s}}^{\dagger}\, \widehat{{\mathcal e}}^{\dagger}},
\label{Q16}
\end{eqnarray}
where $z = r(t) \, e^{i \theta}$ while $\delta_{e} = \omega_{e}\,t$ and $\delta_{s} = \omega_{s} \, t$.
In terms Eqs. (\ref{Q15})--(\ref{Q16}) the results of 
Eqs. (\ref{Q13})--(\ref{Q14}) can be written as:
\begin{eqnarray}
\widehat{s}(t) &=&  \widehat{\Sigma}^{\dagger}(z)\,\,  \widehat{{\mathcal R}}^{\dagger}(\delta) \,\,
\widehat{{\mathcal s}}\,\,\widehat{{\mathcal R}}(\delta)\,\,\widehat{\Sigma}(z),
\label{Q17}\\
\widehat{e}(t) &=&  \widehat{\Sigma}^{\dagger}(z)\,\,  \widehat{{\mathcal R}}^{\dagger}(\delta) \,\,
\widehat{{\mathcal e}}\,\,\widehat{{\mathcal R}}(\delta)\,\,\widehat{\Sigma}(z).
\label{Q18}
\end{eqnarray}
The late-time density operator can then be expressed in terms of $\widehat{{\mathcal R}}(\delta)$ and $\widehat{\Sigma}(z)$:
\begin{equation}
\widehat{\rho}(t,\, t_{in}) = \widehat{{\mathcal R}}(\delta) \, \widehat{\Sigma}(z) \,\, \widehat{\rho}_{s\,e}(t_{in}) \,\,  \widehat{\Sigma}^{\dagger}(z)  \widehat{{\mathcal R}}^{\dagger}(\delta).
\label{Q19}
\end{equation}

\subsection{Symmetries of the Hamiltonian}
As anticipated at the beginning of this discussion, the {\em total} Hamiltonian of Eq. (\ref{Q4})
commutes with the difference between the number operators of the system and of the environment 
\begin{equation}
\widehat{{\mathcal Q}} =\widehat{N}_{e} - \widehat{N}_{s}, \qquad [\,\widehat{H}, \widehat{{\mathcal Q}}\,] =0,
\label{Q20}
\end{equation}
but not with their sum, since 
\begin{equation}
[\widehat{H}, \widehat{N}_{e} + \widehat{N}_{s}] = 2 \widehat{H}_{s\,e} \neq 0.
\end{equation}
This observation implies that both 
the Hamiltonian and  the $\widehat{{\mathcal Q}}$ operators 
admit the same orthonormal and complete set 
of eigenfunctions so that they can be simultaneously diagonalized; as we shall see in section \ref{sec3} this is not the only basis to discuss the irreducible representations (see, in particular, \cite{THIRTYFOUR}) but it is in fact close to the standard 
one where the r\^ole of the Hamiltonian is played by the Casimir operator of the underlying group.
The second remark is that Eq. (\ref{Q20}) justifies the terminology we shall be using
in the forthcoming discussions since the parametric amplification (even in a purely quantum mechanical framework) is in fact equivalent to the production of quanta. This will also be true, a fortiori, in the second quantized perspective that will be further scrutinized in section \ref{sec6}.
 
\renewcommand{\theequation}{3.\arabic{equation}}
\setcounter{equation}{0}
\section{The reduced density operators}
\label{sec3}
 \subsection{The complete density operator}
The late-time density matrices follow directly from Eq. (\ref{Q19}) by requiring, as repeatedly stressed, that the density operators of the system and of the environment are 
initially uncorrelated, as explained prior 
to Eq. (\ref{INT6}). Therefore the initial density matrix of Eq. (\ref{Q19}) shall be 
expressed in the following manner
\begin{equation} 
\widehat{\rho}_{s\,e}(t_{in}) = \widehat{\rho}_{s}(t_{in})\otimes \widehat{\rho}_{e}(t_{in}).
\label{R1}
\end{equation}
At the onset of the evolution the system is in the vacuum (i.e.  $\widehat{\rho}_{s}(t_{in})= |\,0 \rangle \langle 0\,|$) while the density matrix of the environment is a mixture of states with statistical weights provided by the Bose-Einstein (geometric) distribution:
\begin{equation}
\widehat{\rho}_{e}(t_{in}) = \sum_{m= 0}^{\infty}\,\overline{p}_{m}(\overline{n}) \, | \, m\rangle \langle m\,|,
\label{R2}
\end{equation}
where $\overline{n}$ is the averaged multiplicity of the initial state and $\overline{p}_{m}(\overline{n})$ corresponds to the Bose-Einstein distribution\footnote{Since throughout the discussion various statistical weights will progressively appear, the probability distributions associated with the initial density operator of the environment will be supplemented by an overline (e. g. $\overline{p}_{m}$).}
\begin{equation}
\overline{p}_{m}(\overline{n})=\frac{\overline{n}^{m}}{(\overline{n} +1)^{m + 1}}, \quad \sum_{m= 0}^{\infty}\,\overline{p}_{m}(\overline{n}) =1.
\label{R2a}
\end{equation}
In case the environment is in local thermal equilibrium the simplest possibility is that $\overline{n} = (e^{\omega_{e}/T_{e}} -1)^{-1}$. In the presence of a chemical potential the form of $\overline{p}_{m}$ remains the same in terms of the averaged multiplicity but $\overline{n}$ gets modified as $\overline{n} = (e^{(\omega_{e} -\mu_{e})/T_{e}} -1)^{-1}$. This is why, incidentally, it is preferable to express the statistical weights as in Eq. (\ref{R2a}); this form is also practical for further generalizations (see e.g. \cite{THIRTYFIVE} and discussions therein). The complete expression of the total density matrix encompassing the system and the environment can then be written as 
\begin{equation}
\widehat{\rho}(t,\, t_{in}) = \sum_{m=0}^{\infty} \sum_{\ell=0}^{\infty} \sum_{\ell^{\prime}=0}^{\infty} 
{\mathcal A}_{m\ell\ell^{\prime}} | \ell, \ell + m \rangle \langle m + \ell^{\prime}, \ell^{\prime}|, 
\label{R3}
\end{equation}
where ${\mathcal A}_{m,\ell,\ell^{\prime}}$ is given by
\begin{equation}
{\mathcal A}_{m\ell\ell^{\prime}}= P_{m\ell\ell^{\prime}} \sqrt{\binom{m+ \ell}{m} \binom{m+ \ell^{\prime}}{m} },
\label{R4}
\end{equation}
and, besides the two binomial coefficients, $P_{m\ell \ell^{\prime}}$ is
\begin{equation}
P_{m \ell \ell^{\prime}} =  \overline{p}_{m}(\overline{n})\frac{e^{i \alpha (\ell -\ell^{\prime})}}{(\overline{n}^{(q)}+1)^{m+1}} \biggl(\frac{\overline{n}^{(q)}}{\overline{n}^{(q)}+1}\biggr)^{(\ell +\ell^{\prime})/2},
\label{R5}
\end{equation}
with $\alpha = (\theta+\pi - \delta_{e} -\delta_{s})$. Equation (\ref{R5}) gives the complete 
form of the total density operator sometimes referred to as the {\em proper} density 
operator \cite{THIRTY} to distinguish it from the {\em improper} ones possibly obtainable after tracing over 
some of the degrees of freedom belonging either to the system or to the environment.
It is straightforward 
but rather lengthy to show from Eq. (\ref{R3}) that 
\begin{equation}
\mathrm{Tr}\bigl[ \widehat{\rho}(t,\, t_{in}) \bigr] = 1, \qquad 
\mathrm{Tr}\bigl[ \widehat{\rho}^2(t,\, t_{in}) \bigr] \neq 1,
\label{R6}
\end{equation}
as it must happen in the case of a mixed state. In particular 
it can be shown, after simple algebra, that $\mathrm{Tr}\bigl[ \widehat{\rho}^2(t,\, t_{in}) \bigr]$ is given by:
\begin{equation}
 \frac{[\overline{n}^{(q)} +1]^2}{[ 1 + ( 2 + \overline{n})\overline{n}^{(q)}][1 + 2 \overline{n}^{(q)} + \overline{n}(2 + 3 \overline{n}^{(q)})]},
\label{R7}
\end{equation}
implying that, in the limit $\overline{n}^{(q)}\to 0$ (no parametric amplification) 
$\mathrm{Tr}\bigl[ \widehat{\rho}^2(t,\, t_{in})] = 1/( 2 \overline{n} +1)$ 
as it is expected in the case where the statistical weights of the density 
matrix appear in the Bose-Einstein form. Since the present analysis 
involves a number of reduced density operators it is wise 
to check their properties, step by step, after each reduction, as we shall 
be consistently doing in what follows.

\subsection{The reduced density operators}
From the proper density operator of Eq. (\ref{R3}) 
the reduced density matrices are obtained  by tracing $\widehat{\rho}(t,\, t_{in})$ either 
over the degrees of freedom of the system or of the environment (see, e.g. \cite{THIRTY}):
\begin{eqnarray}
&&  \widehat{\rho}_{s}(t,t_{in}) = \mathrm{Tr}_{e}[ \widehat{\rho}(t,\, t_{in})],
\label{TRACEe}\\
&&\widehat{\rho}_{e}(t, t_{in}) = \mathrm{Tr}_{s}[ \widehat{\rho}(t,\, t_{in})],
\label{TRACEs}
\end{eqnarray}
where, as already stressed, $\mathrm{Tr}_{e}$ and $\mathrm{Tr}_{s}$ denote, respectively, the traces 
over the degrees of freedom of the environment and of the system.
The explicit forms of Eqs. (\ref{TRACEe})--(\ref{TRACEs}) are:
\begin{eqnarray}
&& \widehat{\rho}_{e}(t, t_{in})  =  \sum_{\ell=0}^{\infty} \sum_{m=0}^{\infty} p^{(e)}_{m\,\ell}  \, |\,\ell+m \rangle\langle m+ \ell\, |,
\label{EQN}\\
&& \widehat{\rho}_{s}(t, t_{in}) = \sum_{\ell=0}^{\infty} 
p^{(s)}_{\ell}  \, |\,\ell \rangle\langle \ell\, |.
\label{EQO}
\end{eqnarray}
In Eqs. (\ref{EQN})--(\ref{EQO}) the statistical weights connected to the environment and to the system 
are indicated, respectively, by $p^{(e)}_{m\,\ell}$ and by  $p^{(s)}_{\ell}$ and they depend upon $t$ and $t_{in}$.
It is convenient to characterize the final state in terms of the averaged multiplicities of the 
produced quanta since, for the explicit evaluation $\Delta S_{s}$ and $\Delta Q_{e}$, this is what ultimately counts.  In the quantum thermodynamical approach followed here  the statistical weights only depend on the 
averaged multiplicities of the initial state (i.e. $\overline{n}$) and 
of the produced species (i.e. $\overline{n}^{(q)}$). 
In particular the statistical weight associated with the {\em system} is: 
\begin{equation}
 p^{(s)}_{\ell} = \frac{\overline{n}^{(q)\, \ell}\, ( 1 + \overline{n})^{\ell}}{[ 1 + \overline{n}^{(q)}(\overline{n}+1)]^{\ell +1}}.
\label{TRACE3}
\end{equation}
It can be immediately checked from Eq. (\ref{TRACE3}) that the traces of the reduced 
density operator are, as expected, 
\begin{equation}
\sum_{\ell=0}^{\infty}p^{(s)}_{\ell} =1, \quad \mathrm{Tr} \widehat{\rho}_{s} =1,\quad \mathrm{Tr}\widehat{\rho}_{s}^2  \neq 1.
\label{TRACE3a}
\end{equation}
The statistical weights of the {\em environment} 
are instead given by:
\begin{equation}
 p^{(e)}_{\ell\, m}  = \binom{m+ \ell}{m} \frac{ \overline{n}^{m}\, \overline{n}^{(q)\,\,\ell}}{(\overline{n} +1)^{m +1} [\overline{n}^{(q)} +1]^{m + \ell +1}},
\label{TRACE4}
\end{equation}
and again we can verify from Eq. (\ref{TRACE4}) that 
\begin{equation}
\sum_{\ell=0}^{\infty}\,\sum_{m=0}^{\infty}p^{(e)}_{\ell\,m} =1, \quad \mathrm{Tr} \widehat{\rho}_{e} =1, \quad \mathrm{Tr}\widehat{\rho}_{e}^2  \neq 1.
\label{TRACE4a}
\end{equation}
From Eq. (\ref{TRACE4}) we can easily compute the probability generating function $P^{(e)}(s,w)$for the bivariate (discrete) distribution $p^{(e)}_{\ell\, m}$:
\begin{eqnarray}
P^{(e)}(s,w) &=& \sum_{\ell=0}^{\infty} \sum_{m=0}^{\infty} s^{m} \, w^{\ell} \, p^{(e)}_{\ell\, m} 
\nonumber\\
&=& \frac{1}{[ 1 + (1 - s) \overline{n} + (1 - w) \overline{N}]}.
\label{TRACE4b}
\end{eqnarray}
From Eq. (\ref{TRACE4b}) we can immediately appreciate that in the limit $w \to 1$ 
we recover the probability generating function of the Bose-Einstein distribution of the initial state 
(with averaged multiplicity $\overline{n}$). In the limit $s\to 1$ the distribution is still 
of Bose-Einstein type but with averaged multiplicity $\overline{N}= ( 1 + \overline{n}) \overline{n}^{(q)}$.
This explains why the variable $\overline{N}$ appears to be very convenient in the physical discussion.

\subsection{Symmetries of the states}
The results obtained above in this section can be swiftly deduced through a judicious use of the 
symmetries of the underlying quantum states employed in the explicit derivation of the density operators. 
To clarify this suggestion we first note that the free and the interacting Hamiltonians evaluated at the 
onset of the dynamical evolution can be in fact expressed through three operators that shall be conventionally denoted hereunder by
 $\widehat{K}_{\pm}$ and $\widehat{K}_{0}$:
\begin{eqnarray}
\widehat{K}_{+} &=& \widehat{{\mathcal s}}^{\dagger}\,\widehat{{\mathcal e}}^{\dagger}, \qquad  \widehat{K}_{-} = \widehat{{\mathcal s}}\,\widehat{{\mathcal e}},
\nonumber\\
\widehat{K}_{0} &=& (\widehat{{\mathcal s}}^{\dagger} \widehat{{\mathcal s}} + \widehat{{\mathcal e}} \widehat{{\mathcal e}}^{\dagger})/2.
\label{OP1}
\end{eqnarray}
The operators of Eq. (\ref{OP1}) are quadratic in the creation and annihilation operators defined in  Eq. (\ref{Q7a}) 
and satisfy the commutation relations of the $SU(1,1)$ Lie algebra (see, for instance, \cite{THIRTYTHREE} and references therein):
\begin{equation}
[\widehat{K}_{0}, \, \widehat{K}_{\pm}] = \pm \widehat{K}_{\pm}, \qquad [\widehat{K}_{+}, \,\widehat{K}_{-}] = -2\, \widehat{K}_{0}.
\label{OP2}
\end{equation}
Furthermore in terms of $\widehat{K}_{\pm}$ and $\widehat{K}_{0}$  the Casimir operator becomes
\begin{equation}
\widehat{{\mathcal C}}= \widehat{K}_{0}^2 - (\widehat{K}_{+}\,\widehat{K}_{-} 
+ \widehat{K}_{-}\,\widehat{K}_{+})/2.
\label{OP3}
\end{equation}
Thanks to the commutation relations of Eq. (\ref{OP2}) the expression of $\widehat{{\mathcal C}}$ can also 
be written as 
\begin{equation}
\widehat{{\mathcal C}}= \widehat{K}_{0}(\widehat{K}_{0} -1)  - \widehat{K}_{+}\,\widehat{K}_{-}. 
\end{equation}
Since, by definition, the Casimir operator commutes with all the generators of the group, a possible 
basis for the irreducible representations of $SU(1,1)$ corresponds to the Bargmann choice \cite{THIRTYFOUR}
where $\widehat{{\mathcal C}}$ and $\widehat{K}_{0}$ are simultaneously diagonalized.
In the context of the present problem a more convenient basis for the irrreducible representation of $SU(1,1)$ is\footnote{The connection between the Bargmann basis and the one of Eq. (\ref{OP4}) 
is given by $m_{e} + m_{s} = (2 \overline{m} -1)$ and $m_{e} - m_{s} = (2 k -1)$. In the literature $\overline{m}$ is often denoted by $m$; the notation $\overline{m}$ is preferred here to avoid potential confusions with various summation indices that may appear throughout the discussions. }:
\begin{equation}
| m_{s},\, m_{e} \rangle = \frac{({\mathcal s}^{\dagger})^{m_{s}}}{\sqrt{m_{s}!}} \, \frac{({\mathcal e}^{\dagger})^{m_{e}}}{\sqrt{m_{e}!}} | 0_{s},\,0_{e}\rangle.
\label{OP4}
\end{equation}
The density matrix encompassing 
the system and the environment at $t_{fin}$ can be ultimately related to the state 
\begin{equation}
|\delta\, z\rangle = \widehat{{\mathcal R}}(\delta) \, \widehat{\Sigma}_{+}(z)\widehat{\Sigma}_{0}(z)\widehat{\Sigma}_{-}(z) | m_{s},\, m_{e} \rangle,
\label{OP7}
\end{equation}
defined in the Schr\"odinger description. 
For an explicit expressions of  $|\delta\, z\rangle$ we can use the Baker-Campbell-Hausdorff decomposition for the $SU(1,1)$ group \cite{THIRTYSIX} (see also \cite{THIRTYSEVEN,THIRTYEIGHT,THIRTYNINE}) implying that the operator $\widehat{\Sigma}(z)$ is factorized as the product of the exponentials of the group generators i.e. 
\begin{equation}
 \widehat{\Sigma}(z) =  \widehat{\Sigma}_{+}(z)\,  \widehat{\Sigma}_{0}(z)\,  \widehat{\Sigma}_{-}(z),
\label{OP7a}
\end{equation}
where $\widehat{\Sigma}_{\pm}(z)$ and $\widehat{\Sigma}_{0}(z)$ are:
\begin{eqnarray}
\widehat{\Sigma}_{\pm}(z) &=& \exp{[ \mp e^{ \pm i \vartheta} \tanh{r}\widehat{K}_{\pm}]}, 
\nonumber\\
\widehat{\Sigma}_{0}(z) &=& \exp{[- 2 \ln{(\cosh{r})} \widehat{K}_{0}]}.
\label{OP7b}
\end{eqnarray}
Thanks to Eqs. (\ref{OP7a})--(\ref{OP7b}) we can then obtain from Eq. (\ref{OP7}):
\begin{eqnarray}
|\delta\, z\rangle &=& \sum_{j = 0}^{j_{max}} \sum_{\ell=0}^{\infty} 
  \frac{e^{i \, \gamma( j, \ell)}\,  (\cosh{r})^{2 \, j}\,(\tanh{r})^{j + \ell}}{(\cosh{r})^{ m_{s} +m_{e}+1} }  
\nonumber\\
&\times& {\mathcal M}(j, \, \ell)\, |m_{s} - \ell + j,\, m_{e} -\ell + j\rangle,
\label{OP8}
\end{eqnarray}
where $j_{max} = \mathrm{Min}[m_{s},\, m_{e}]$ while 
$\gamma( j, \ell) = [ (\delta_{s} +\delta_{e} - \theta)(j -\ell) + \pi\ell - 
(\delta_{s} \,m_{s} + \delta_{e} \, m_{e})]$. The explicit  expression of ${\mathcal M}(j,\, \ell)$ is finally given by:
\begin{equation}
{\mathcal M}(j,\, \ell)= \frac{\sqrt{m_{s}!\, m_{e}!} \sqrt{(m_{s} - j + \ell)!\, (m_{e} - j + \ell)!}}{j !\, \ell!\, (m_{s} - j)!\, (m_{e} - j)!}.
\label{OP9}
\end{equation}
Thanks to Eqs. (\ref{OP8})--(\ref{OP9}) the relevant initial states of the system and of the environment can be 
constructed and evolved in time. As repeatedly stressed in the present and in the previous sections,  we are chiefly interested in the case where the system is in the vacuum while the environment 
is in a mixed state. In the language of Eqs. (\ref{OP8})--(\ref{OP9}) the system is initially 
in the vacuum provided $m_{s} =0$; this means that, by definition, also $j_{max} \to 0$ and the 
sum appearing in Eq. (\ref{OP8}) reduces to a single contribution, i.e. $j =0$. In practice the final 
state will be given by Eqs. (\ref{OP8})--(\ref{OP9}) evaluated for $m_{s} = j = 0$ and the result
becomes 
\begin{eqnarray}
|\Psi_{m_{e}}(t,\,t_{in})\rangle &=& \sum_{\ell =0}^{\infty} \frac{(\tanh{r})^{\ell} \, e^{i (\alpha\ell - \delta m_{e})}}{(\cosh{r})^{m_{e} +1}}
\nonumber\\
&\times& \sqrt{\frac{(m_{e} + \ell)!}{m_{e} !\, \ell !}} \, |\ell, \, m_{e} + \ell \rangle.
\label{OP10}
\end{eqnarray}
If Eq. (\ref{OP7a}) is directly applied to the initial state 
state  $|0_{s}\, m_{e}\rangle = |0_{s}\rangle \, |m_{e}\rangle$ we obtain 
\begin{equation}
|\Psi_{m_{e}}(t,\,t_{in})\rangle = \widehat{{\mathcal R}}(\delta) \, \widehat{\Sigma}_{+}(z)\, \widehat{\Sigma}_{0}(z)  |0_{s}\, m_{e}\rangle,
\label{OP10a}
\end{equation}
which coincides with the result of Eq. (\ref{OP10}) deduced from Eq. (\ref{OP9}) in the limit $m_{s} \to 0$ and $j\to 0$. According to the previous considerations (see, in particular, Eqs. (\ref{INT6}) and (\ref{R2a})) from the initial thermal state for the environment the proper density 
operator becomes 
\begin{equation}
\widehat{\rho}(t,\,t_{in}) = \sum_{m =0}^{\infty} \overline{p}_{m} \, |\Psi_{m}(t,\,t_{in})\rangle \, \langle  \Psi_{m}(t,\,t_{in})|,
\label{OP11}
\end{equation}
where, as before, $\overline{p}_{m}= p_{m}^{(in)}$ is the Bose-Einstein (geometric) distribution associated with the initial density matrix of the thermal environment; note that, for  simplicity, we redefined 
the eigenvalue as $m_{e} \to m$. It also follows from Eqs. (\ref{OP10})--(\ref{OP11}) that the total density matrix coincides with the result of Eq. (\ref{R3}). 
 
As we close this section, it is appropriate to comment on the representations of $SU(1,1)$ by means of operators bilinear in the boson creation and annihilation operators $\widehat{a}^{\dagger}$ and $\widehat{a}$ satisfying canonical commutation relations $[ \widehat{a}, \widehat{a}^{\dagger}] =1$. This realization is possible because 
the $SU(1,1)$ and the $Sp(2,R)$ groups are isomorphic\footnote{The operators 
$L_{+} = \widehat{a}^{\dagger\, 2}/2$, $L_{-} = \widehat{a}^{2}/2$ and $L_{0}= (\widehat{a}^{\dagger} \widehat{a} + 
\widehat{a} \widehat{a}^{\dagger})/4$ actually close the $SU(1,1)$ algebra. In this case the eigenvalues of
the Casimir operator correspond either to $k =1/4$ or to $k =3/4$. The basis of the unitary representation is, in this case, $| n \rangle = (n!)^{-1/2} (\widehat{a}^{\dagger})^{n} | 0 \rangle$; for $n$ even the unitary representation 
corresponds to $k =1/4$ while for $n$ odd the unitary representation corresponds to $3/4$.} \cite{THIRTYTHREE}. This observation is at the heart of the first applications of two-photon coherent states in quantum optics \cite{THIRTYFOURa,THIRTYFOURb,THIRTYFOURc}.
There are however various physical differences between the problem analyzed here and the analog application possibly based on two-photon optics. The two-mode structure is actually essential to analyze the interaction between the system and the environment from a quantum thermodynamical viewpoint.

\renewcommand{\theequation}{4.\arabic{equation}}
\setcounter{equation}{0}
\section{The heat flow and the entropy}
\label{sec4} 
\subsection{The heat transferred to the environment}
The heat transferred to the environment and the increment of the entropy of the system shall now be carefully evaluated since 
both results are essential for the derivation of the entropy bounds 
that are quantitatively examined in section \ref{sec5}.
Recalling Eqs. (\ref{INT10})--(\ref{INT11}) the heat flowing to the environment 
is deduced from the variation of the Hamiltonian between the final and the initial states. 
In the Heisenberg description $\Delta Q_{e}$ is  
\begin{equation}
\Delta Q_{e} = \langle \mathrm{th} |\widehat{H}_{e}(t_{fin}) |\mathrm{th}\rangle - \langle \mathrm{th} |\widehat{H}_{e}(t_{in})  |\mathrm{th}\rangle,
\label{QQ1}
\end{equation}
where $| \mathrm{th} \rangle$ indicates the initial thermal state of the environment 
characterized by the averaged thermal multiplicity $\overline{n}$.
The results obtained from Eq. (\ref{QQ1}) must also follow from the density matrix $\widehat{\rho}_{e}$ of Eq. (\ref{TRACEs}) coming from the trace over the degrees of freedom of the system (see also Eq. (\ref{INT11})) and discussion therein:
 \begin{equation}
\Delta Q_{e} = \mathrm{Tr}\biggl\{ \widehat{H}_{e} \biggl[ \widehat{\rho}_{e}(t_{fin}) - \widehat{\rho}_{e}(t_{in}) \biggr]\biggr\},
\label{QQ2}
\end{equation}
where now $\widehat{H}_{e} = \widehat{H}_{e}(t_{in})$ is evaluated in the Schr\"odinger description.
For the consistency of the whole approach it is interesting to check both derivations; 
in particular, if we start from Eq. (\ref{QQ1}), we obtain 
\begin{equation}
\Delta Q_{e} = \omega_{e} \langle \mathrm{th}| \widehat{N}_{e}(t_{fin}) | \mathrm{th} \rangle  - 
\omega_{e} \langle \mathrm{th}| \widehat{N}_{e}(t_{in}) | \mathrm{th} \rangle.
\label{QQ3}
\end{equation}
Since the explicit form of $\widehat{N}_{e}(t_{fin}) = \widehat{e}^{\dagger}(t_{fin})  \widehat{e}(t_{fin})$
can be computed with the help of Eq. (\ref{Q14}), Eq. (\ref{QQ3}) becomes:
\begin{eqnarray}
\Delta Q_{e} &=& \omega_{e}\bigl( \overline{n} \cosh^2{r} + \sinh^2{r} - \overline{n}\bigr)
\nonumber\\
&=& \omega_{e} \, \overline{n}^{(q)} \, (\overline{n} +1).
\label{QQ4}
\end{eqnarray}
Equation (\ref{QQ4}) confirms the strategy already adopted in Eqs. (\ref{EQN})--(\ref{EQO}) (see also Eq. (\ref{TRACE4b})) for the quantitative discussion of the forthcoming results. Indeed the averaged multiplicity of the created quanta (i.e. $\overline{n}^{(q)} = \sinh^2{r}$) together 
with the thermal multiplicity $\overline{n}$ can be used, in their various combinations, as the pivotal variables 
of the problem. We also point out that from now on $r$ indicates for the sake of simplicity,  the value of $q(t)$ 
integrated between $t_{in}$ and $t_{fin}$: 
\begin{equation}
r= r(t_{fin}, t_{in}) = \int_{t_{in}}^{t_{fin}} q(t) \, dt.
\label{QQ4a}
\end{equation}
Equations (\ref{QQ4})--(\ref{QQ4a}) clarify once more why, in Eqs. (\ref{TRACE3}) and (\ref{TRACE4}), we found 
useful to express the statistical weights directly in terms of $\overline{n}$ and $\overline{n}^{(q)}$: $\overline{n}$ defines the onset of the evolution whereas $\overline{n}^{(q)}$ characterizes the final asymptotic stages of the dynamics. Equation (\ref{QQ4}) coincides with 
the results obtained from the reduced density matrix; more specifically along this perspective we have\footnote{The result of Eq. (\ref{QQ5}) simply expresses the general truism that quantum fluctuations are relevant at low temperatures since the expectation value
of the Hamiltonian $\widehat{H}_{e}$ over a thermal state characterized by a Bose-Einstein probability distribution is exactly $\langle\mathrm{th}| \widehat{H}_{e} |\mathrm{th} \rangle = \omega_{e} (\overline{n} +1/2)$. Furthermore given that $\overline{n}= (e^{\omega_{e}/T} -1)^{-1}$, $\langle\mathrm{th}| \widehat{H}_{e} |\mathrm{th} \rangle$ goes as $T$ for 
 $\omega_{e} \ll T$  while, in the opposite limit (i.e. $\omega_{e} \gg T$), $\langle\mathrm{th}| \widehat{H}_{e} |\mathrm{th} \rangle$ coincides in practice with the ground state energy.}
\begin{eqnarray}
&&\mathrm{Tr}\bigl[ \widehat{H}_{e}  \widehat{\rho}_{e}(t_{in}) \bigr] = \omega_{e} \bigl[ \overline{n} +1/2\bigr],
\label{QQ5}\\
&& \mathrm{Tr}\bigl[ \widehat{H}_{e}  \widehat{\rho}_{e}(t_{fin}) \bigr] = \omega_{e} \bigl[1/2 + 
\overline{n} + \overline{n}^{(q)}( \overline{n} +1)\bigr],
\label{QQ6}
\end{eqnarray}
where Eq. (\ref{QQ5}) is immediately obvious while Eq. (\ref{QQ6}) is a direct consequence 
of Eqs. (\ref{EQN}) and (\ref{TRACE4}). If we now subtract Eq. (\ref{QQ5}) from Eq. (\ref{QQ6}) we obtain, 
again, the result of Eq. (\ref{QQ4}). This is because, in any case, 
the contribution of the ground state cancels when deriving $\Delta Q_{e}$; 
the factor $1/2$ appearing in Eqs. (\ref{INT14})--(\ref{INT15}) is therefore not essential 
for the result of Eq. (\ref{QQ4}) so that we could have defined the original free 
Hamiltonians by renormalizing, as occasionally done, the zero-point energy. 

\subsection{The variation of the entropy of the system}
From Eqs. (\ref{INT10}) we can now compute the variation of the entropy of the 
system, namely
\begin{equation}
\Delta\, S_{s}= S[ \widehat{\rho}_{s}(t_{fin})] - S[ \widehat{\rho}_{s}(t_{in})],
\label{SS1}
\end{equation}
 where, as usual, $S[\widehat{\rho}] = - \mathrm{Tr}[\widehat{\rho} \, \ln{\widehat{\rho}}]$ 
 is the von Neumann entropy which is the one customarily employed in the 
 quantum mechanical derivation of the Landauer's bound. From Eq. (\ref{SS1}) we can immediately
 note that $S[ \widehat{\rho}_{s}^{(in)}]=0$ since the initial (pure) state of the 
 system coincides with the vacuum. It also follows from Eqs. (\ref{EQO}) and (\ref{TRACE3}) that Eq. (\ref{SS1})
 can be rewritten as:
 \begin{eqnarray}
 \Delta\, S_{s} &=& - \sum_{\ell=0}^{\infty} p_{\ell}^{(s)} \ln{ p_{\ell}^{(s)}}
 \nonumber\\
 &=& (\overline{N} +1) \ln{(\overline{N}+1)} - \overline{N} \ln{\overline{N}},
 \label{SS2}
 \end{eqnarray}
 where $p_{\ell}^{(s)}$ has been deduced in Eq. (\ref{TRACE3}). It is convenient to express 
 Eq. (\ref{SS2}) directly in terms of the global variable 
\begin{equation}
\overline{N} = \overline{n}^{(q)} \, (\overline{n} +1) = (\overline{n} +1)\,\sinh^2{r}.
\label{SS3}
\end{equation}
In the limit $r \to 0$ we have $\overline{N}\to 0$ and $\Delta\, S_{s} \to 0$ so that
the entropy of the system does not change in this case since the averaged multiplicity 
of the produced quanta vanishes (i.e. $\overline{n}^{(q)} \to 0$); recalling Eq. (\ref{QQ4a}) this 
happens, in practice, when $t_{fin} \to t_{in}$ and the system remains in its initial stage of evolution
without appreciable interaction with the environment.  Therefore the notation of Eq. (\ref{SS3}) is physically meaningful since 
$\overline{N}$ ultimately accounts for the total averaged multiplicity of the produced quanta.

\renewcommand{\theequation}{5.\arabic{equation}}
\setcounter{equation}{0}
\section{The physical bounds}
\label{sec5}
\subsection{The entropy variation and the heat flow}
To clarify the nature of the bounds derived in this investigation the first useful step is to compute the ratio between the 
increment of the entropy {\em of the system}  (i.e. $T \Delta S_{s}$) and the heat transferred {\em to the environment} (i.e. $\Delta Q_{e}$). After constructing the ratio $T\, \Delta S_{s}/\Delta Q_{e}$ we shall eventually discuss if it is generically smaller than $1$, as suggested in Eq. (\ref{INT2}). Bearing in mind the stenographic notation spelled out in Eq. (\ref{SS3}), the wanted expression is immediately obtained from Eqs. (\ref{QQ4}) and (\ref{SS2}); the final  
result solely depends upon $\overline{N}$ and upon $(T/\omega_{e})$:
\begin{equation}
T \frac{\Delta S_{s}}{\Delta Q_{e}}= \biggl(\frac{T}{\omega_{e}}\biggr)\biggl[\frac{\ln{\overline{N}}}{\overline{N}} + \biggl(1 + \frac{1}{\overline{N}}\biggr) \ln{\biggl(1 + \frac{1}{\overline{N}}\biggr)}\biggr].
\label{NPB4}
\end{equation}
\begin{figure}[ht!]
\begin{centering}
\includegraphics[width=7cm,height=7cm]{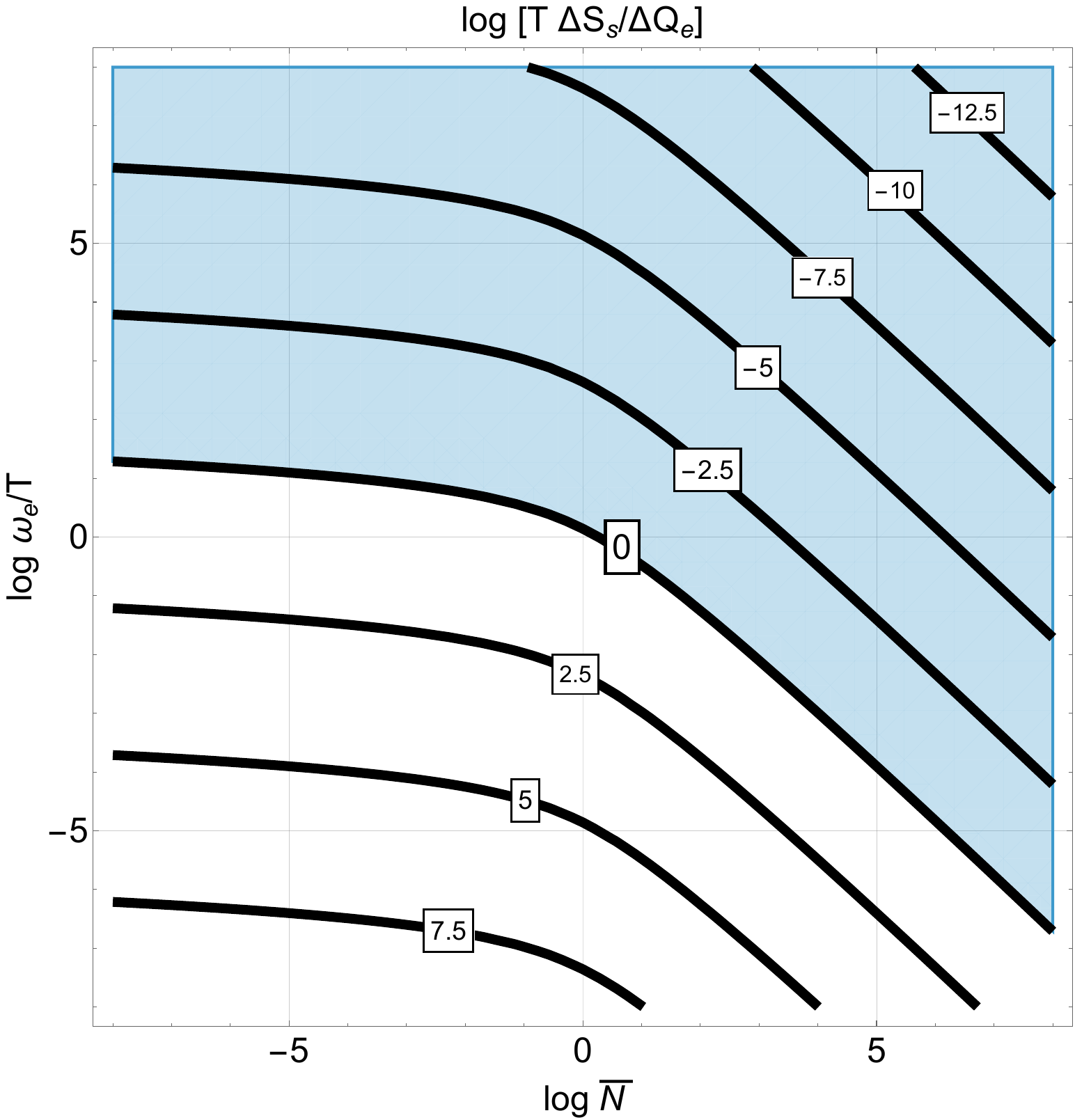}
\par\end{centering}
\caption{\label{FIGURE1} In this and in the following plots of this section the shaded areas illustrates the regions where $T  \Delta S_{s} < \Delta Q_{e}$. 
Common logarithms are employed on both axes. The labels are constant along each contour and correspond to the common logarithms of $T  \Delta S_{s}/\Delta Q_{e}$ computed from the right-hand side of Eq. (\ref{NPB4}).  In this plot the bounds stemming from Eq. (\ref{NPB4}) are studied in the plane $(\overline{N}, \, \omega_{e}/T)$. The white areas where the bound is not satisfied correspond to the region of large thermal multiplicities and comparatively negligible quantum amplification (i.e. $\overline{n}^{(q)} = {\mathcal O}(1)$ or smaller). }
\end{figure}
A simple algebraic inspection of Eq. (\ref{NPB4}) suggests that $T\, \Delta S_{s} < \Delta Q_{e}$ provided the conditions of the parametric amplification are met and this means, in particular, that $\overline{N} >1$. The shaded areas in Fig. \ref{FIGURE1} and in the forthcoming plots correspond to the regions where $T \Delta S_{s} \leq \Delta Q_{e}$. According to the results of Fig. \ref{FIGURE1} the entropy growth connected to $\Delta S_{s}\geq 0$ is indeed bounded by $\Delta Q_{e}$, as long as the quantum parametric amplification is operational (i.e. $\overline{N}> 1$). 

A complementary perspective might however suggest that since $\overline{N}= \overline{n}^{(q)} ( \overline{n} + 1)$ the condition $\overline{N}>1$ is realized when $\overline{n}^{(q)}> 1$ while the averaged thermal multiplicity $\overline{n}$ could be both larger and smaller than $1$. Therefore, to avoid confusions and to clarify the matter even further we are going illustrate the right hand side of Eq. (\ref{NPB4}) not only in the plane $(\overline{N}, \, \omega_{e}/T)$ (as already done in Fig. \ref{FIGURE1}) but also in terms of the other variables that already appeared in the previous considerations 
of sections \ref{sec3} and \ref{sec4}. The second point we wish to examine is what happens when $\overline{N} <1$. In this 
regime the bounds derived here are likely to be violated simply because the system did not evolve from the initial vacuum state. In the following 
two subsections we shall separately examine the regimes $\overline{N}> 1$ and $\overline{N}<1$. 
\begin{figure}[ht!]
\begin{centering}
\includegraphics[width=7cm,height=7cm]{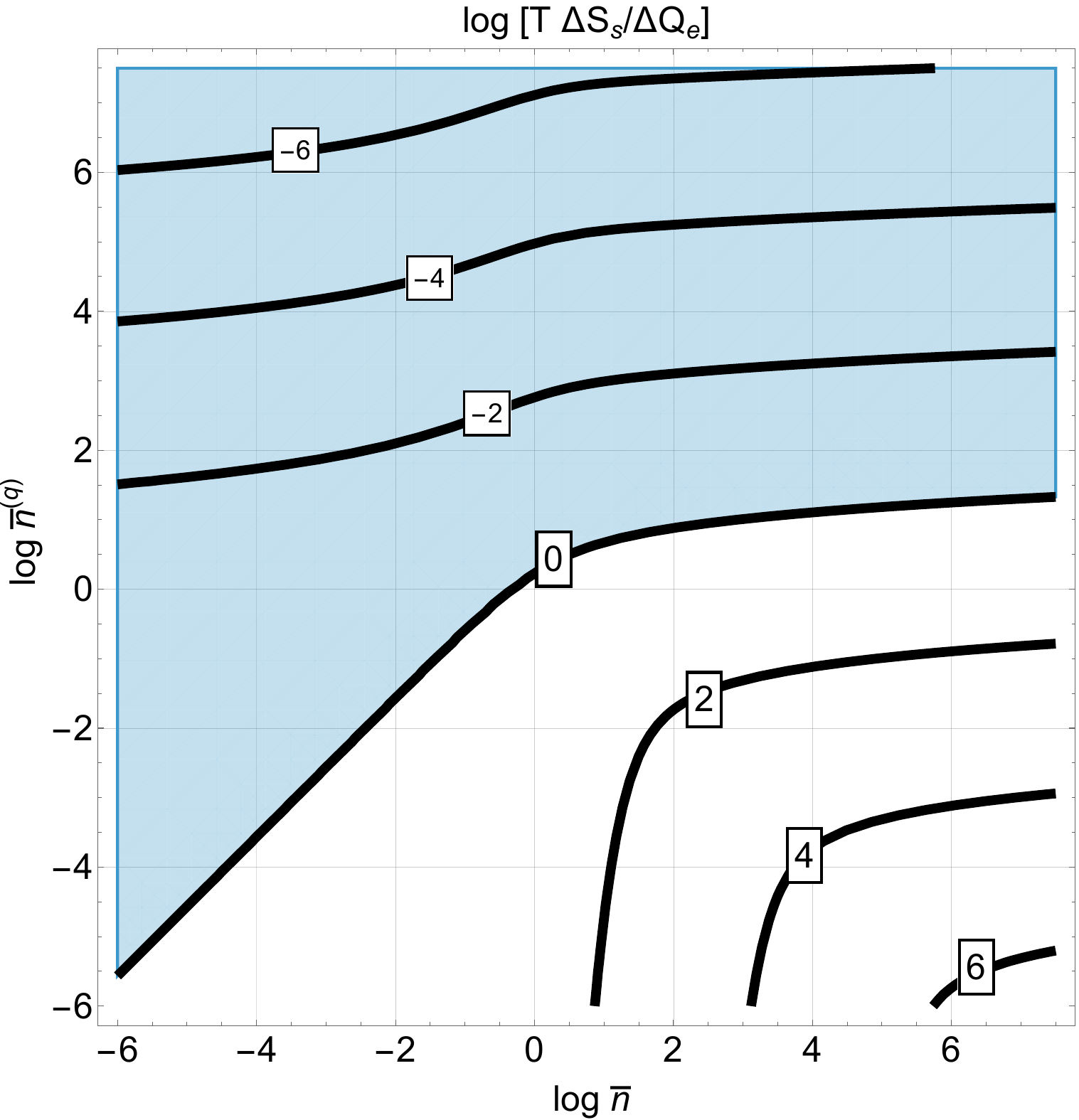}
\par\end{centering}
\caption{\label{FIGURE2} As in in Fig. \ref{FIGURE1} the labels appearing on the various contours 
correspond to the common logarithms of $T  \Delta S_{s}/\Delta Q_{e}$. The bound $T\, \Delta S_{s} < \Delta Q_{e}$ is here scrutinized in the plane defined by the averaged multiplicities of the initial thermal state (i.e. $\overline{n}$) and of the produced quanta  (i.e. $\overline{n}^{(q)}$). The bound is always verified, in practice, when $\overline{n}^{(q)} > 1$ and in spite of the value of $\overline{n}$ except for the region of small multiplicities where the limits derived from Eq. (\ref{NPB6}) are effective (i.e. $\overline{n} \,\, \overline{n}^{(q)} < e^{\overline{n}^{(q)}}$).}
\end{figure}
\subsection{The regime $\overline{N} > 1$}
When $\overline{N} > 1$  the right hand side of Eq. (\ref{NPB4}) can be expanded in powers of $1/\overline{N}$ and if 
only the leading terms of the expansion in $(1/\overline{N})$ are kept, Eq. (\ref{NPB4}) becomes:
\begin{equation}
T \frac{\Delta S_{s}}{\Delta Q_{e}} = \frac{1 + \ln{\overline{N}}}{\overline{N} \, \ln{(1 + 1/\overline{n})}},
\label{NPB5}
\end{equation}
where $T/\omega_{e}$ has been traded for the averaged thermal multiplicity $\overline{n}$.  As long as  $\overline{n}\gg 1$ the logarithm in the denominator can be further expanded in the limit $(1/\overline{n}) <1$. Moreover, in the same limit (i.e. $\overline{n}\gg 1$) Eq. (\ref{SS3}) implies that $\overline{N} = \overline{n}^{(q)} \, \overline{n}$. Putting together the two previous observations, the following expression is finally deduced from Eq. (\ref{NPB5}): 
\begin{equation}
T \frac{\Delta S_{s}}{\Delta Q_{e}} = \frac{1 + \ln{(\overline{n}^{(q)} \, \overline{n})}}{\overline{n}^{(q)}}.
\label{NPB6}
\end{equation}
The result of Eq. (\ref{NPB6}) implies that $T \,\Delta S_{s} < \Delta Q_{e}$ provided $\overline{n}$ 
and $\overline{n}^{(q)}$ satisfy the approximate inequality 
$\overline{n} < e^{\overline{n}^{(q)} -1}/\overline{n}^{(q)}$. Given that the  amplification is operational only when the quanta are effectively produced  (i.e. $\overline{n}^{(q)} > 1$), the obtained condition gets even simpler, i.e.  $\overline{n} \, \overline{n}^{(q)} < e^{\overline{n}^{(q)}}$. Consequently, when $\overline{n}^{(q)} \gg 1$ the condition on $\overline{n}$ is always verified, in practice, due to the 
largeness of the exponential factor. Once more, from Fig. \ref{FIGURE2} we can appreciate that the white region for $\overline{n}>1$ corresponds to the values of $\overline{n}^{(q)}$ that remain sufficiently small in agreement with the condition $\overline{n} \, \overline{n}^{(q)} < e^{\overline{n}^{(q)}}$.
\begin{figure}[ht!]
\begin{centering}
\includegraphics[width=6.8cm,height=6.8cm]{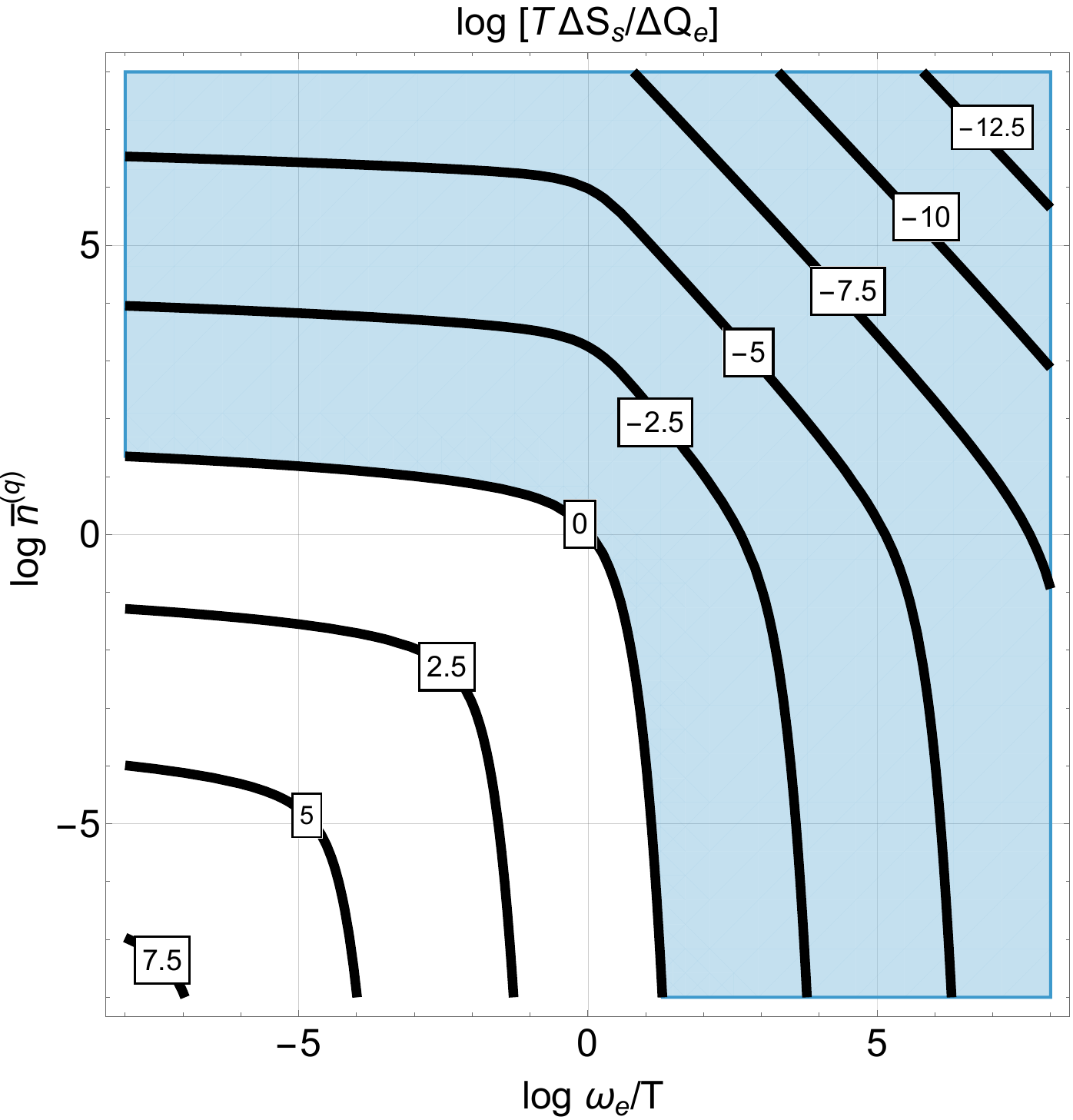}
\par\end{centering}
\caption{\label{FIGURE3} We illustrate the contours of constant  $T  \Delta S_{s}/\Delta Q_{e}$ in the plane 
$(\omega_{e}/T,\, \overline{n}^{(q)})$. As usual, the labels on the various contours indicate the common logarithm 
of $T  \Delta S_{s}/\Delta Q_{e}$. From this plot it is particularly clear that in the region $\omega_{e}/T \gg 1$ (i.e. $\overline{n} \ll 1$) and $\overline{n}^{(q)} >1$  the condition $T \Delta S_{s} < \Delta Q_{e}$ is always verified, as expected from Eq. (\ref{NPB7}). }
\end{figure}

Always under the hypothesis that $\overline{N} >1$, we finally examine the case where $\overline{n} \ll 1$. 
This means that, in practice, the total averaged multiplicity is entirely generated quantum mechanically, i.e. $\overline{N} \simeq \overline{n}^{(q)}$. When
$\overline{n} \ll 1$ we also have that $T/\omega_{e} \ll 1$ and Eq. (\ref{NPB5}) becomes:
\begin{equation}
T \frac{\Delta S_{s}}{\Delta Q_{e}} = \frac{1 + \ln{\overline{N}}}{\overline{N}} \biggl(\frac{T}{\omega_{e}}\biggr).
\label{NPB7}
\end{equation}
We see from Eq. (\ref{NPB7}) that $T \Delta S_{s} < \Delta Q_{e}$ since the condition
$\overline{N} \geq 1 + \ln{\overline{N}}$ is always verified and $T/\omega_{e} \ll 1$.
For the sake of completeness in Fig. \ref{FIGURE3} we then illustrate the bound in the plane $(\omega_{e}/T,\, \overline{n}^{(q)})$; as in the previous plots the labels appearing in the contours indicate the 
common logarithm of $T  \Delta S_{s}/\Delta Q_{e}$. It is also clear from Fig. \ref{FIGURE3} 
that the bound is even verified in the regime $\overline{n}^{(q)}\ll 1$ provided $\omega_{e}/T\gg 1$.

\subsection{The regime $\overline{N} < 1$}
Recalling the explicit expression of Eq. (\ref{SS3}, when $\overline{N} < 1$ the quantum amplification is effectively absent.
This means, in practice, that the entropy of the system does not decrease but does not increase either. 
Therefore this regime we expect that the bounds are violated simply because 
the system and the environment remain close to their initial states where the corresponding density matrices are uncorrelated. 
Let us then go back to Eq. (\ref{NPB4}) and note that, without approximations,  the general expression can be rephrased as
\begin{equation}
T \frac{\Delta S_{s}}{\Delta Q_{e}} = \frac{(\overline{N} +1)\ln{(\overline{N}+ 1)} - \overline{N} \ln{\overline{N}}}{\overline{N} \ln{(1 + 1/\overline{n})}}.
\label{NPC1}
\end{equation} 
If the numerator at the right hand side of Eq. (\ref{NPC1}) is expanded in the limit $\overline{N} <1$ the resulting 
expression becomes
\begin{equation}
T \frac{\Delta S_{s}}{\Delta Q_{e}} = \frac{1 - \ln{\overline{N}}}{\ln{(1 + 1/\overline{n})}}, \qquad \overline{N}<1.
\label{NPC2}
\end{equation}
Again, because of Eq. (\ref{SS3}) the condition $\overline{N}<1$ must correspond to $\overline{n}^{(q)}< 1$. 
\begin{figure}[ht!]
\begin{centering}
\includegraphics[width=7cm,height=7cm]{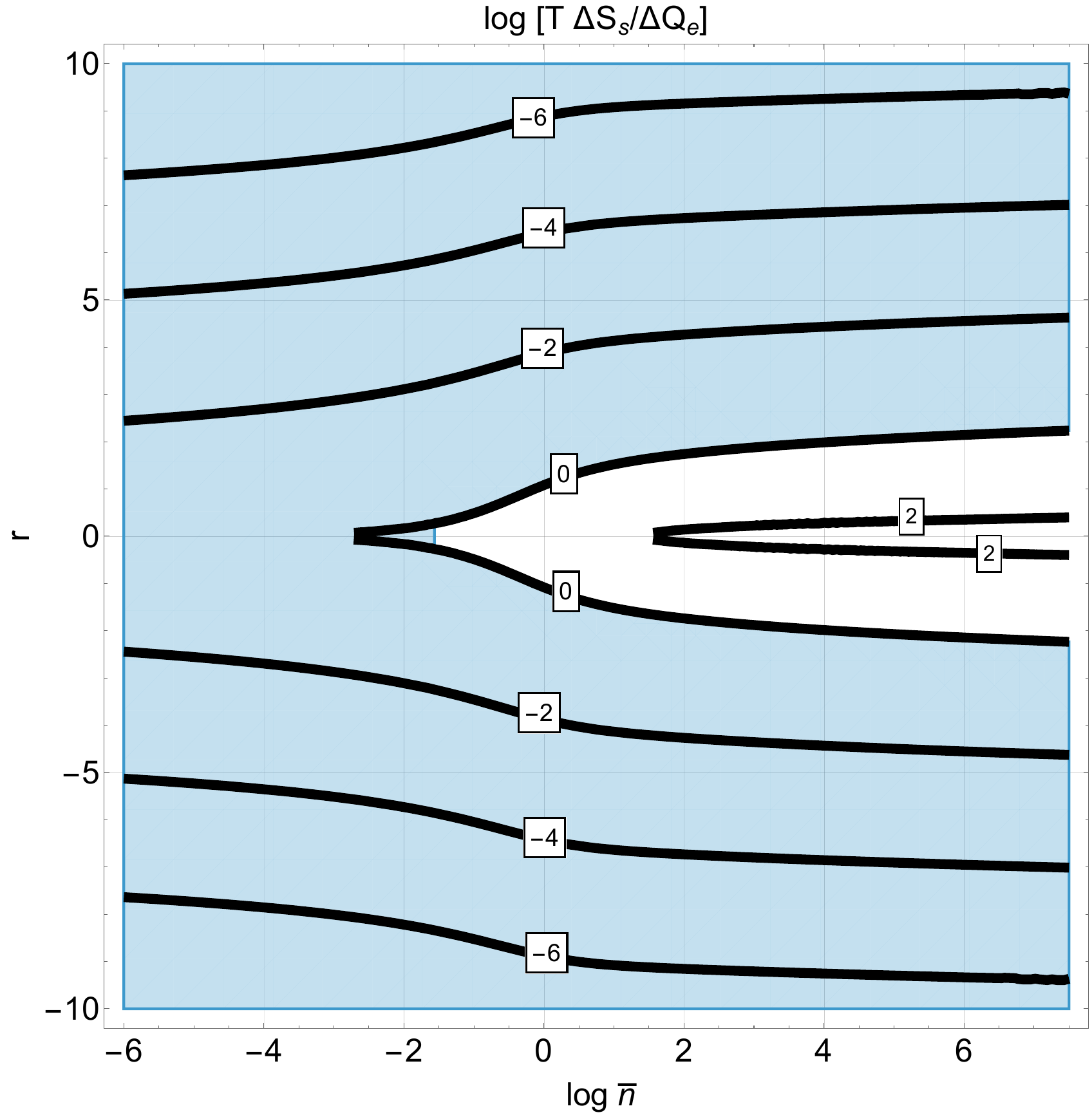}
\par\end{centering}
\caption{\label{FIGURE4} We illustrate the contours of constant  $T  \Delta S_{s}/\Delta Q_{e}$ in the plane 
$(\overline{n},\, r)$. We recall that since $\overline{n}^{(q)}= \sinh^2{r}$, the values of $r$ can be both positive and negative. The 
$r$-parametrization is particularly suitable for the analysis of the region $\overline{n}^{(q)}<1$. As usual the labels on the various contours indicate the common logarithm of $T  \Delta S_{s}/\Delta Q_{e}$. }
\end{figure}
To investigate this portion of the parameter space, it is very convenient to recall that the averaged multiplicity of the produced quanta is,  by definition, $\overline{n}^{(q)} = \sinh^2{r}$ (see Eqs. (\ref{QQ4}) and (\ref{SS3}) and discussions thereafter). When $\overline{n}^{(q)} < 1$ the parameter $r$ can be both negative and positive provided $|r| \leq 0.88$. 
In this regime the quanta are not produced 
but a certain class of operators (the so called quadrature operators \cite{THIRTEEN,FOURTEEN}) may fluctuate either above or below the quantum noise. What matters, for the present ends, is however 
that even in the regime $| r| \to 0$ the entropy variation is 
bounded by the heat transfer except for a tiny slice centred around $| r | \to 0$. 
As usual the shaded area in Fig. \ref{FIGURE4}  corresponds to $T \Delta S_{s} \leq \Delta Q_{e}$ in the 
plane $(\overline{n},\, r)$. However, for $\overline{n} \gg 1$
and $\overline{n}^{(q)} < 1$ both the system and the environment 
remain in their (uncorrelated) initial state as in the absence of any dynamical evolution.
If we go back to Eq. (\ref{NPC2}) we can first consider the case 
$\overline{n} < 1$ so that our expression ultimately becomes\footnote{To derive Eq. (\ref{NPC3}) we noted, as usual, that from Eq. (\ref{SS3}) $\overline{N}  = \overline{n}^{(q)}( \overline{n} +1) \simeq  \overline{n}^{(q)}$,
while $1/\overline{n} > 1$. }:
\begin{equation}
T \frac{\Delta S_{s}}{\Delta Q_{e}} = \frac{1 - \ln{\overline{n}^{(q)}}}{\ln{ (1/\overline{n}})}, \qquad \overline{n} <1.
\label{NPC3}
\end{equation}
Equation (\ref{NPC3}) implies that $T \Delta S_{s} < \Delta Q_{e}$ as long as 
$\ln{(\overline{n}^{(q)}/\overline{n})} > 1$ which also means $e \, \overline{n} < \overline{n}^{(q)} < 1$.
We may finally consider the opposite case, i.e. $\overline{n} \gg 1$. Now the total final multiplicity becomes $\overline{N} = \overline{n}^{(q)} (\overline{n} +1) \simeq \overline{n}^{(q)} \,\overline{n}$ and the minuteness of $\overline{n}^{(q)}$ should counterbalance the largeness 
of $\overline{n}$ so that, overall, $\overline{N} = \overline{n}^{(q)} \,\overline{n} < 1$. From Eq. (\ref{NPC1}) we have 
that the analog of Eq. (\ref{NPC3}) is
\begin{equation}
T \frac{\Delta S_{s}}{\Delta Q_{e}} = \overline{n}[ 1 - \ln{(\overline{n}^{(q)} \, \overline{n})}], \qquad \overline{n} > 1.
\label{NPC4}
\end{equation}
\begin{figure}[ht!]
\begin{centering}
\includegraphics[width=7cm,height=7cm]{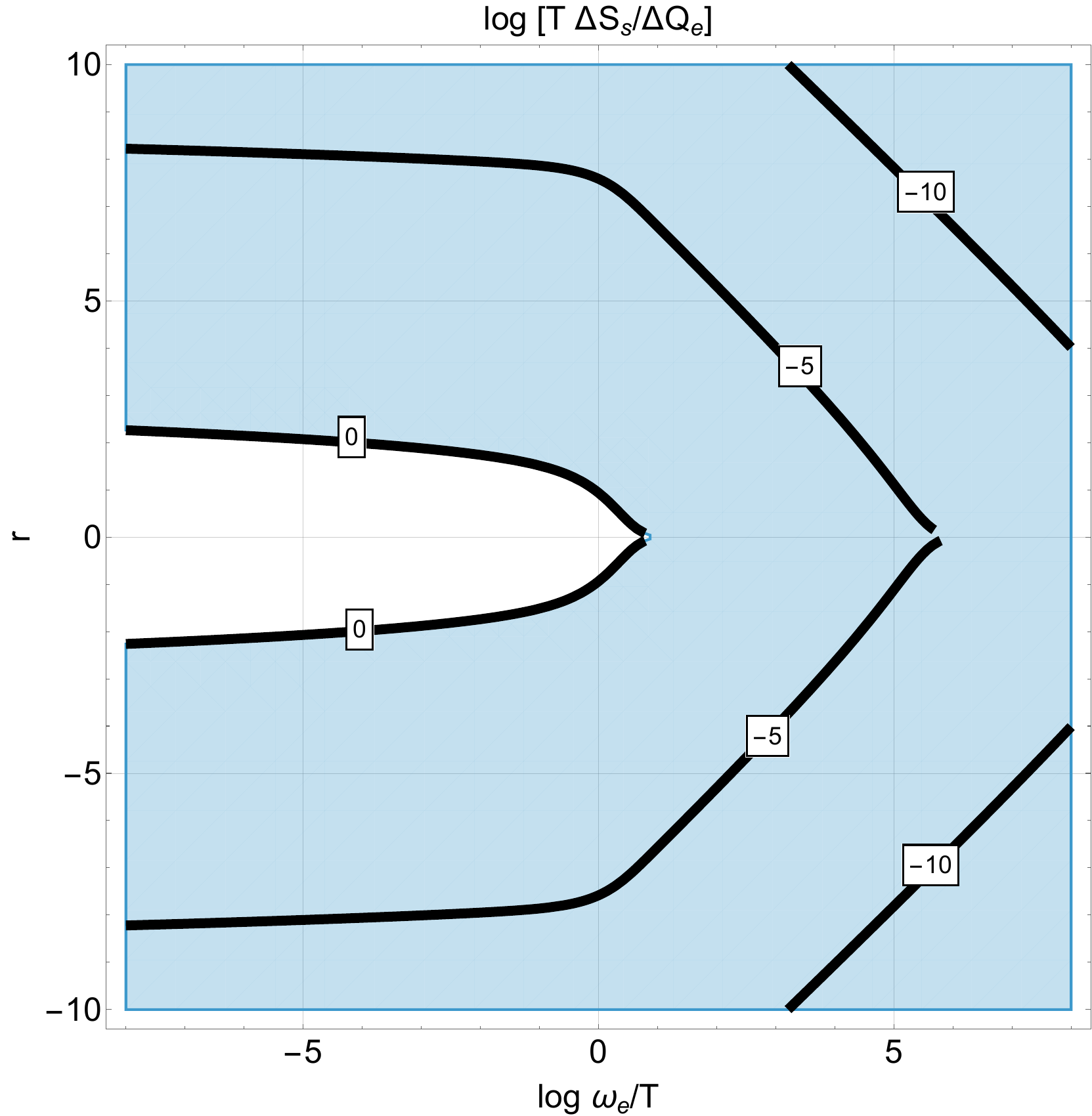}
\par\end{centering}
\caption{\label{FIGURE5} We illustrate the same region discussed in Fig. \ref{FIGURE4} but in the 
plane $(\omega_{e}/T,\, r)$. As usual the shaded area correspond to the region where the 
variation of the entropy of the system is bounded by the heat transferred to the environment.}
\end{figure}
According to Eq. (\ref{NPC4})  $\ln{(\overline{n}^{(q)}\, \overline{n})} > 1- 1/\overline{n}$ which also 
implies, for $\overline{n}\gg 1$, that $\overline{n}^{(q)}\, \overline{n}> e$. The latter condition 
is {\em incompatible} with the hypothesis $\overline{n}^{(q)}\, \overline{n} < 1$ 
that has been assumed in Eq. (\ref{NPC4}). The result (\ref{NPC4}) is expected since, in the limit $r\to 0$ 
and $\overline{n} \gg 1$, the environment and the system remain, in practice, in their (uncorrelated) 
initial states and entropy is not produced. 

In Fig. \ref{FIGURE5} we illustrate, for completeness, the bound on the entropy variation 
in the plane $(\omega_{e}/T,\, r)$. By looking together at Figs. \ref{FIGURE4} and \ref{FIGURE5}
we see that, as anticipated, the bound   $T\, \Delta S_{s}<\Delta Q_{e}$ is always except for a slice 
corresponding to $|r |\to 0$ and $\overline{n} \gg1$. In this region we have that, in practice, 
$\widehat{H}_{s\,e} \to 0$ since the coupling vanishes asymptotically. 
\begin{figure}[ht!]
\begin{centering}
\includegraphics[width=7cm,height=7cm]{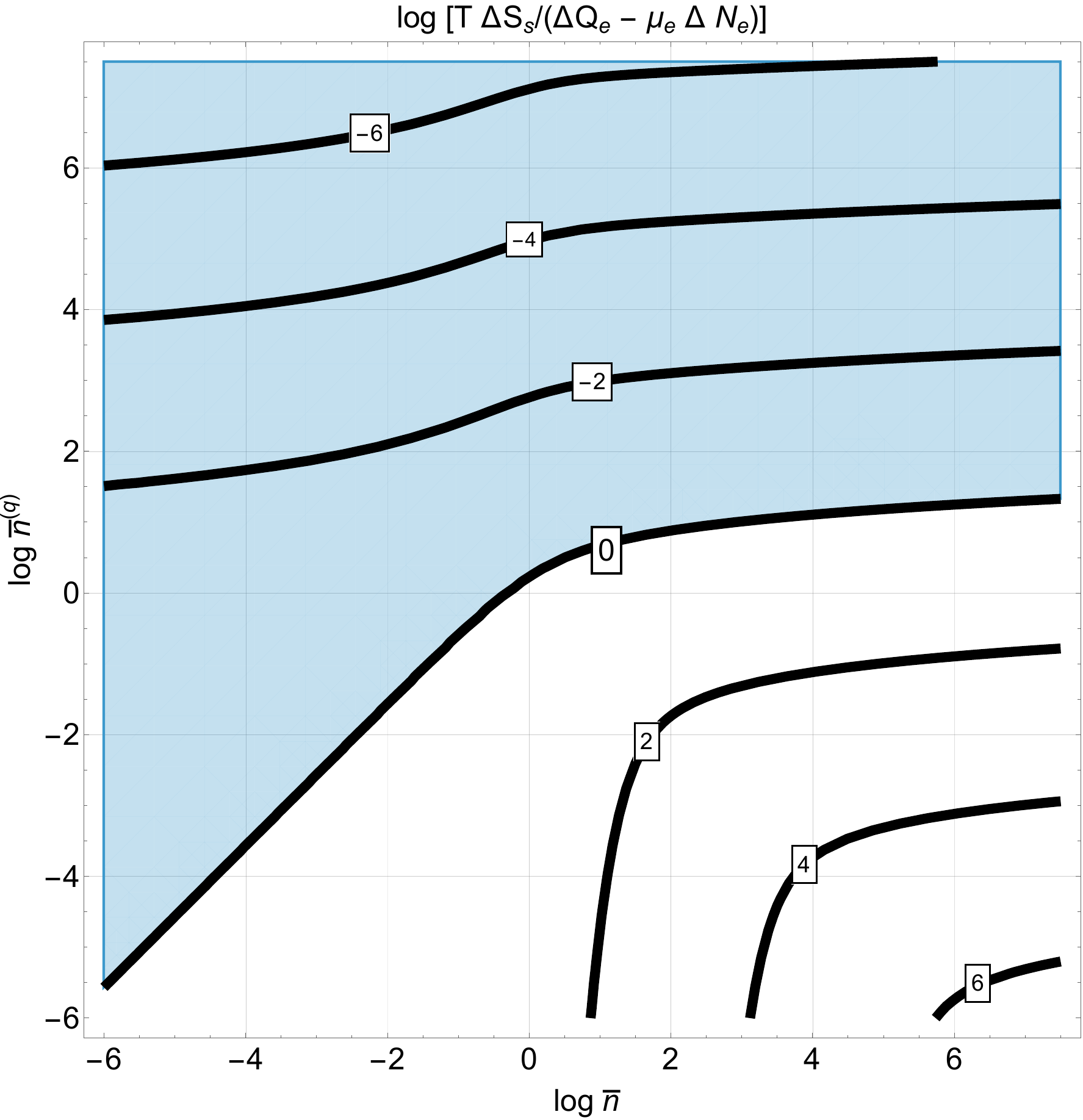}
\par\end{centering}
\caption{\label{FIGURE6} We illustrate the conditions (\ref{NPC8})--(\ref{NPC9}). The shaded area correspond to the region where $T \, \Delta S_{s} \leq \Delta Q_{e} - \mu_{e} \Delta N_{e}$. This plot coincides with Fig. \ref{FIGURE2}; this is because, in the 
presence of the chemical potential, $T/(\omega_{e} - \mu_{e}) = \ln{(1 + 1/\overline{n})}$ so that, ultimately, the condition (\ref{NPC9}) corresponds exactly to the one already deduced, for instance, in Eq. (\ref{NPC1}).}
\end{figure}

\subsection{The issue of the chemical potential}
The quantum mechanical formulation of the Landauer's conjecture does not involve the presence 
of a chemical potential and this choice determines
 the density matrix of the initial state where the system is in the vacuum whereas the environment is in a thermal state characterized by a temperature $T_{e}$ (see Eqs. (\ref{R1})--(\ref{R2}) and discussions thereafter). If
the environment contains a chemical potential the statistical weights of Eq. (\ref{R2a}) are formally the same (i.e. $\overline{p}_{n} = \overline{n}^{n}/(\overline{n} +1)^{n +1}$) but the averaged multiplicity depends both on $T_{e}$ and $\mu_{e}$, i.e.
\begin{equation}
\overline{n} = \frac{1}{ e^{(\omega_{e} - \mu_{e})/T_{e}} -1}.
\label{NPC5}
\end{equation}
Th presence of a chemical potential is plausible as long as the flow of energy associated 
with $\Delta Q_{e}$ is complemented by a flow of quanta measured 
by $\Delta N_{e}$. We remind, in this respect, that the total Hamiltonian of our 
system commutes with the difference $\widehat{N}_{e} - \widehat{N}_{s}$, but does not commute with $\widehat{N} = \widehat{N}_{s} + \widehat{N}_{e}$
\begin{equation}
[\widehat{H}, \widehat{N}] \neq 0, \quad [\widehat{H}, \widehat{N}_{e} - \widehat{N}_{s}] = 0,
\label{NPC6}
\end{equation}
see also Eq. (\ref{Q20}) and discussions thereafter.
If a chemical potential is present the bound on the entropy variation of the system 
can be phrased as 
\begin{equation}
 T \, \Delta S_{s} \leq \Delta Q_{e} - \mu_{e} \Delta N_{e},
 \label{NPC7}
 \end{equation}
 where, consistently with the previous notations, we set $T= T_{e}$. 
  As in the case $\mu_{e} \to 0$ the bound of Eq. (\ref{NPC7}) 
 is always verified under the same conditions already deduced above this section. 
 To reach this conclusion the simplest route is to observe 
 that the ratio between $T \, \Delta S_{s}$ and $(\Delta Q_{e} - \mu_{e} \Delta N_{e})$
 is still given by:
 \begin{equation}
 \biggl(\frac{T}{\omega_{e} -\mu_{e}}\biggr)
 \frac{(\overline{N}+1) \ln{(\overline{N}+1)} - \overline{N} \ln{\overline{N}}}{\overline{N}}.
 \label{NPC8}
 \end{equation}
 But from Eq. (\ref{NPC5}) we can always trade $T/(\omega_{e} - \mu_{e})$ for 
 $1/\ln{(1 + 1/\overline{n})}$ and  this means that the bound (\ref{NPC7}) is satisfied 
 provided
 \begin{equation}
  \frac{(\overline{N}+1) \ln{(\overline{N}+1)} - \overline{N} \ln{\overline{N}}}{\overline{N}\,\ln{(1 + 1/\overline{n})}} <1. 
 \label{NPC9}
 \end{equation} 
 The condition (\ref{NPC9}) can be studied in the plane $(\overline{n}, \, \overline{n}^{(q)})$
and the result of this analysis coincides exactly with Fig. \ref{FIGURE2}. The reason 
is that the condition $T \Delta S_{s} \leq \Delta Q_{e}$ deduced from 
Eq. (\ref{NPB4}) in the plane $(\overline{n}^{(q)}, \, \overline{n})$ takes the 
same form of Eq. (\ref{NPC9}). This is because the addition of the chemical potential in the 
bound is always compensated by the modified form of the averaged thermal multiplicity 
of Eq. (\ref{NPC5}). In Fig. \ref{FIGURE6} the conditions 
(\ref{NPC7}) and (\ref{NPC8})--(\ref{NPC9}) are illustrated in the plane $(\overline{n}, \, \overline{n}^{(q)})$.

\subsection{A complementary comment}
As we close this discussion we wish to comment on the possible direct 
verifications of the bounds suggested here. Although it is 
 difficult to endorse specific experimental platforms,
we find nonetheless useful to remind few of possibilities that have been repeatedly 
explored in the last forty years, namely the backward \cite{THIRTYNINEa} and 
the forward \cite{THIRTYNINEb} four-wave mixing as well as the parametric down 
conversion \cite{THIRTYNINEc}. In spite of the obvious differences,
the quantum thermodynamical considerations discussed here are 
closely related to the formation of squeezed states of light (see e.g. \cite{THIRTEEN,FOURTEEN})
that have been originally scrutinized, within complementary frameworks, forty years ago
\cite{THIRTYNINEa,THIRTYNINEb,THIRTYNINEc}. The two modes associated with the system and the environment can therefore be empirically identified with two appropriate modes of the electromagnetic field in a cavity or in an optical fiber.  Although in the last forty years an enormous technical progress has been made in the generation of squeezed light (see e.g. \cite{THIRTYNINEd} for a relatively recent review), the experimental platforms of nonlinear crystals, optical fibers and atomic ensembles 
exploited in Refs. \cite{THIRTYNINEa,THIRTYNINEb,THIRTYNINEc} essentially 
coincide with the ones used today for generating strongly squeezed light. 
It seems therefore plausible that similar platforms might be 
effectively used for the empirical analysis of the entropic bounds 
derived in this investigation. Of course we are only expressing here a theoretical viewpoint 
(based on the similarities between the symmetries of the underlying problems) rather than 
a concrete empirical suggestion.

\renewcommand{\theequation}{6.\arabic{equation}}
\setcounter{equation}{0}
\section{The perspective of field theory}
\label{sec6}
The field theoretical perspective
rests on the same physical premises of the quantum parametric amplification
 and the final forms of the entropy bounds are in fact very similar  
with the crucial difference that the averaged multiplicities, the final states 
and the density operators exhibit an explicit momentum dependence. 
For the sake of concreteness we start by examining the following parametrization 
of the action of a single scalar degree of freedom $\psi(\vec{x},\tau)$ in four space-time dimensions:
\begin{equation}
S = \frac{1}{2} \int d^3x \, \int \,d\tau \,v^2(\tau) \biggl[\partial_{\tau} \psi \partial_{\tau} \psi - \partial_{\ell} \psi \partial^{\ell} \psi\biggr],
\label{FT1}
\end{equation}
where $\tau$ denotes the conformal time coordinate which is widely employed in curved backgrounds and in cosmological applications \cite{FORTY}; in the case of a conformally flat background metric of Friedmann-Robertson-Walker type we have that $a(\tau) d \tau = d t$. The time dependence is encoded in the function $v(\tau)$ that ultimately 
 determines the specific form of the pump fields entering the effective Hamiltonian. Since different time 
parametrizations could be employed \cite{FIFTEEN,SIXTEEN} (see also \cite{SEVENTEEN,EIGHTEEN}), the introduction of $\tau$ is not strictly essential but it is convenient also in the light of potential applications. In Eq. (\ref{FT1}) $\partial_{\ell}$ denotes, as usual, the spatial 
gradients and the related indices are Euclidean.  The action (\ref{FT1}) can be reduced to its canonical form after redefining the fields as 
$\mu(\vec{x},\tau) = v(\tau) \, \psi(\vec{x},\tau)$ so that, eventually, the Lagrangian density 
becomes 
\begin{eqnarray}
{\mathcal L}(\vec{x},\tau) = \frac{1}{2}\biggl[ (\partial_{\tau}\mu)^2 + {\mathcal F} \mu^2 
- 2 \,{\mathcal F}\, \mu\, \partial_{\tau}\mu - (\vec{\nabla}\mu)^2 \bigg],
\label{FT2}
\end{eqnarray}
where ${\mathcal F} = v^{\prime}/v$ and the prime indicates hereunder a derivation 
with respect to the conformal time coordinate $\tau$. The canonical momentum deduced from Eq. (\ref{FT2}) is given by 
$ \pi = (\partial_{\tau} \mu - {\mathcal F} \mu)$ and from Eq. (\ref{FT1}) the Hamiltonian becomes
\begin{equation}
H(\tau) = \frac{1}{2} \int d^{3} x \biggl[ \pi^2 + 2 \,{\mathcal F}\, \mu\,\pi + (\vec{\nabla}\mu)^2 \biggr],
\label{FT3}
\end{equation}
where $\vec{\nabla}$ denotes, as usual, the gradient in three spatial dimensions.  
Depending on the actual meaning of $v(\tau)$, ${\mathcal F}$ and $\psi$
both the action (\ref{FT1}) and the Hamiltonian (\ref{FT3}) may correspond to diverse physical situations potentially involving the  production of particles in curved backgrounds. For instance, if
$\psi(\vec{x},\tau)$ coincides with a minimally coupled scalar field in a (conformally flat) 
Friedmann-Robertson-Walker background, $v(\tau)$ equals the scale 
factor $a(\tau)$ expressed in the conformal time parametrization. In the case of the evolution of curvature inhomogeneities 
$\psi(\vec{x},\tau) = {\mathcal R}(\vec{x},\tau)$ where $ {\mathcal R}(\vec{x},\tau)$ indicates the curvature inhomogeneity on comoving orthogonal hypersurfaces. In this second situation $v(\tau) = a(\tau) \varphi^{\prime}/{\mathcal H}$ where ${\mathcal H} = a^{\prime}/a$ and $\varphi$ is the inflaton field \cite{FORTY}. Last but not least, as we shall see, Eq. (\ref{FT3}) may also describe the evolution 
of a single tensor polarization in a Friedmann-Robertson-Walker background \cite{NINE} 
(see also Refs. \cite{TWENTYTWOc,TWENTYTWOd} and \cite{FORTYTHREE}).

\subsection{The quantum description of the problem}
Following the standard approach the classical fields of Eq. (\ref{FT3}) 
are now promoted to the status of quantum (Hermitian) 
operators obeying canonical commutation relations at equal times. This means that 
$\mu \to \widehat{\mu}$, $\pi \to \widehat{\pi}$ and 
\begin{equation}
[\widehat{\mu}(\vec{x}, \tau), \widehat{\pi}(\vec{y},\tau) ] = i \, \delta^{(3)}(\vec{x} - \vec{y})
\label{FT4}
\end{equation}
where $\widehat{\mu}^{\dagger} = \widehat{\mu}$ and $\widehat{\pi}^{\dagger} = \widehat{\pi}$.
The Hamiltonian operator deduced from Eq. (\ref{FT3}) becomes: 
\begin{equation}
\widehat{H}(\tau) = \frac{1}{2} \int d^{3} x \biggl[ \widehat{\pi}^2 + {\mathcal F}( \widehat{\mu}\,\widehat{\pi} + \widehat{\pi}\,\widehat{\mu}) + (\vec{\nabla}\widehat{\mu})^2 \biggr].
\label{FT5}
\end{equation}
Both the field operators and the Hamiltonian can be represented in Fourier space; in particular we can write  
\begin{equation}
\widehat{\mu}_{\vec{k}}(\tau) = \frac{1}{(2 \pi)^{3/2}} \int e^{i \vec{k}\cdot\vec{x}} \, \widehat{\mu}(\vec{x},\tau)\, d^{3} k= \widehat{\mu}_{-\vec{k}}^{\dagger}(\tau).
\label{FT6}
\end{equation}
By simply replacing in Eq. (\ref{FT6}) $\widehat{\mu}(\vec{x},\tau)$ and $\widehat{\mu}_{\vec{k}}$ with $\widehat{\pi}(\vec{x},\tau)$ and $\widehat{\pi}_{\vec{k}}$ we obtain the Fourier decomposition for the canonical momenta which are also Hermitian (i.e. $\widehat{\pi}_{\vec{k}} = \widehat{\pi}_{-\vec{k}}^{\dagger}$). From Eq. (\ref{FT4}) the commutation relations in Fourier space become $[\widehat{\mu}_{\vec{k}}(\tau), \widehat{\pi}_{\vec{p}}(\tau)] = i \, \delta^{(3)}(\vec{k} + \vec{p})$. The creation and the annihilation operators connected to the canonical field operators are introduced, as usual, from:
\begin{eqnarray}
&& \widehat{\mu}_{\vec{k}}= \bigl(\widehat{a}_{\vec{k}} + \widehat{a}_{-\vec{k}}^{\dagger}\bigr)/\sqrt{2 k},
\nonumber\\
&& \widehat{\pi}_{\vec{k}} = - i \, \sqrt{k/2} \bigl(\widehat{a}_{\vec{k}} - \widehat{a}_{-\vec{k}}^{\dagger}\bigr).
\label{FT7}
\end{eqnarray}
where, for continuous $\vec{k}$-modes, $[\widehat{a}_{\vec{k}}, \widehat{a}_{\vec{p}}] = \delta^{(3)}(\vec{k} - \vec{p})$. The explicit form of the Hamiltonian operator can then 
be expressed either in terms of $\widehat{\mu}_{\vec{k}}$ and $ \widehat{\pi}_{\vec{k}}$
or by means of the creation and annihilation operators. Between these two options the latter 
is more relevant than the former if we want to clarify the analogy with the quantum theory of parametric 
amplification. Thanks to Eqs. (\ref{FT6})--(\ref{FT7}), the Hamiltonian of Eq. (\ref{FT5}) becomes:
\begin{eqnarray}
\widehat{H}(\tau) &=& \frac{1}{2} \int d^{3}k \, k\, \bigl[ \widehat{a}_{\vec{k}}^{\dagger} \,\widehat{a}_{\vec{k}} + 
\widehat{a}_{-\vec{k}}\,\widehat{a}_{-\vec{k}}^{\dagger}\bigr]
\nonumber\\
&+&\frac{1}{2} \int d^{3}k \bigl[ \sigma(\tau)\,
\widehat{a}_{\vec{k}}^{\dagger} \,\widehat{a}_{-\vec{k}}^{\dagger} +\sigma^{\ast}(\tau) \,\widehat{a}_{\vec{k}} \widehat{a}_{-\vec{k}}\bigr],
\label{FT8}
\end{eqnarray}
where $\sigma= i {\mathcal F}$. Since Eq. (\ref{FT8}) exhibits the two-mode structure 
discussed in the previous sections, the idea is now  to associate half of the spectrum 
of the modes with the system (for instance the modes with momentum $\vec{k}$) while the 
remaining modes (the ones with momentum $- \vec{k}$) will belong to the environment.

\subsection{The two-mode Hamiltonian and its symmetries}
From a practical viewpoint the discussion of Eq. (\ref{FT8}) is more concise if the field operators have a discrete spectrum as it happens in a box of volume $V$; in this way the explicit expressions 
of the quantum states are more concise, as it is well known in the context of quantum optical applications \cite{THIRTEEN,FOURTEEN} (see also \cite{THIRTYFIVE}). As it is well known the integrals over the wavenumbers appearing in Eq. (\ref{FT8}) are replaced sums and,  more 
precisely, when the volume of the box becomes 
very large the connection between the two is:
\begin{equation}
\sum_{\vec{k}} \to V\, \int d^{3}k/(2\pi)^{3}, \quad \mu_{\vec{k}} \to \sqrt{(2\pi)^3/V}\, \mu_{\vec{k}}.
\nonumber
\end{equation}
Similar changes apply to the creation and annihilation operators and to their commutation relations 
that now become $[\widehat{a}_{\vec{k}}, \, \widehat{a}_{\vec{k}}^{\,\dagger}] = \delta_{\vec{k},\, \vec{p}}$ 
where now $\delta_{\vec{k},\, \vec{p}}$ indicates the Kr\"oneker symbol replacing the Dirac delta distribution 
of the continuous mode representation. Equation (\ref{FT8}) can be rewritten as $\widehat{H}(\tau) = \sum_{\vec{k}} \, \widehat{H}_{\vec{k}}(\tau)$ where the two-mode Hamiltonian $\widehat{H}_{\vec{k}}$ is now given by:
\begin{eqnarray}
\widehat{H}_{\vec{k}}(\tau) = \omega_{k} \biggl( \widehat{s}_{\vec{k}}^{\,\,\dagger} \,\widehat{s}_{\vec{k}} + 
\widehat{e}_{\vec{k}}\, \widehat{e}_{\vec{k}}^{\,\,\dagger} \biggr) + g\,\widehat{s}_{\vec{k}}^{\,\,\dagger}\,\widehat{e}_{\vec{k}}^{\,\,\dagger} + g^{\ast} \,\widehat{s}_{\vec{k}}\,\widehat{e}_{\vec{k}}.
\label{FT9}
\end{eqnarray}
We note that, for consistency with the previous notations, we introduced $\omega_{k}^2 = k/2$ and $g = \sigma/2$;  furthermore, as previously suggested, Eq. (\ref{FT9}) follows from Eq. (\ref{FT8}) by identifying the $\widehat{a}_{\vec{k}}$ with the oscillators of the system (i.e. $ \widehat{a}_{\vec{k}} \to \widehat{s}_{\vec{k}} $) 
and $\widehat{a}_{-\vec{k}}$ with the ones of the environment (i.e. $ \widehat{a}_{-\vec{k}} \to \widehat{e}_{\vec{k}} $) \cite{NINE}.  All the results obtained in the previous sections can be generalized to the field theory context since, besides the momentum dependence, Eqs. (\ref{INT14})--(\ref{INT15}) and (\ref{Q3}) are analogous to Eq. (\ref{FT9}). So, for instance, the evolution of the field operators follows from the total Hamiltonian $\widehat{H} = \sum_{\vec{k}} \widehat{H}_{\vec{k}}$ 
\begin{equation}
\partial_{\tau} \widehat{s}_{\vec{p}} = i [\widehat{H}, \widehat{s}_{\vec{p}}], \quad
\partial_{\tau} \widehat{e}_{\vec{p}} = i [\widehat{H}, \widehat{e}_{\vec{p}}].
\label{FT10}
\end{equation}
Recalling the explicit form of the commutation relations (i.e. $[\widehat{s}_{\vec{k}}, \widehat{s}_{\vec{p}}^{\dagger}] =  [\widehat{e}_{\vec{k}}, \widehat{e}_{\vec{p}}^{\dagger}] = \delta_{\vec{k},\,\vec{p}}$ 
and  $[\widehat{s}_{\vec{k}}, \widehat{e}_{\vec{k}}] =0$) the following pair of equations can be deduced from Eq. (\ref{FT10}) 
\begin{equation}
\widehat{s}_{\vec{p}}^{\,\,\prime} = - i \, \omega_{p} \widehat{s}_{\vec{p}} - i g\,\widehat{e}_{\vec{p}}^{\,\dagger}, \quad
\widehat{e}_{\vec{p}}^{\,\,\prime} = - i \, \omega_{p} \widehat{e}_{\vec{p}} - i g\,\widehat{s}_{\vec{p}}^{\,\dagger}.
\label{FT11}
\end{equation}
Although  the time dependence of $g(\tau)$ affects Eq. (\ref{FT11}) (and ultimately 
determines the $k$-dependence of the averaged multiplicities)
what matters for the present ends is the general form of the density 
operators of the system and of the environment as a function of the final multiplicities, as 
already described in the quantum mechanical case. We can therefore express the solution of Eq. (\ref{FT11}) as
\begin{eqnarray}
\widehat{s}_{\vec{k}}(\tau) &=& u_{k}(\tau, \tau_{in})\,\,\widehat{{\mathcal s}}_{\vec{k}} - v_{k}(\tau, \tau_{in}) \,\,\widehat{{\mathcal e}}_{\vec{k}}^{\,\,\dagger},
\nonumber\\
\widehat{e}_{\vec{k}}^{\,\,\dagger}(\tau) &=& u_{k}^{\ast}(\tau, \tau_{in})\,\, \widehat{{\mathcal e}}_{\vec{k}}^{\,\,\dagger} - v_{k}^{\ast}(\tau, \tau_{in}) \,\,\widehat{{\mathcal s}}_{\vec{k}},
\label{FT12}
\end{eqnarray}
where, as in Eq. (\ref{Q7a}), we adopted the shorthand notations 
$\widehat{{\mathcal s}}_{\vec{k}} = \widehat{s}_{\vec{k}}(\tau_{in})$ and 
$\widehat{{\mathcal e}}_{\vec{k}} = \widehat{e}_{\vec{k}}(\tau_{in})$. 

The evolution of $u_{k}(\tau, \tau_{in})$ and $v_{k}(\tau, \tau_{in})$ mirrors the one already analyzed in Eqs. (\ref{Q8})--(\ref{Q9})
\begin{eqnarray}
u_{k}^{\,\prime} &=& - i \omega_{k} \, u_{k} + i g\, v_{k}^{\ast}, 
\nonumber\\
v_{k}^{\,\prime} &=& - i \omega_{k} \, v_{k} + i g\, u_{k}^{\ast}.
\label{FT13}
\end{eqnarray}
For a continuous and differentiable 
pump field  the two (complex) functions $u_{k}(\tau, \tau_{in})$ 
and $v_{k}(\tau, \tau_{in}) $ are anyway subjected to the condition $|u_{k}(\tau,\tau_{in})|^2 -
|v_{k}(\tau,\tau_{in})|^2 =1$; this means that $u_{k}(\tau,\tau_{in})$ and $v_{k}(\tau,\tau_{in})$ 
can be parametrized in terms of $3$ real functions conventionally denoted by 
$\delta_{k}(\tau)$, $\theta_{k}(\tau)$ and $r_{k}(\tau)$ in full analogy with the quantum mechanical situation discussed before:
\begin{eqnarray}
&& u_{k}(\tau,\tau_{in}) = e^{- i \delta_{k}} \cosh{r_{k}},
\nonumber\\
&& v_{k}(\tau,\tau_{in}) = e^{- i (\delta_{k} - \theta_{k})} \sinh{r_{k}}.
\label{FT14}
\end{eqnarray}
The evolution of $\delta_{k}(\tau)$, $\theta_{k}(\tau)$ and $r_{k}(\tau)$ can be 
deduced from Eq. (\ref{FT13}) once the expression of $g(\tau)$ is fixed\footnote{So for instance 
in the case of a minimally coupled scalar field $g(\tau) = i {\mathcal H}/2$ and Eq. (\ref{FT14}), once 
inserted into Eq. (\ref{FT13}), implies a set of evolution equations for $\delta_{k}(\tau)$, $\theta_{k}(\tau)$ and $r_{k}(\tau)$. Within the present notations these equations read 
$r_{k}^{\prime} = {\mathcal H} \cos{(2 \delta_{k} - \theta_{k})}$,
$\delta_{k}^{\prime} = \omega_{k} - {\mathcal H}\, \tanh{r_{k}}\, \sin{(2 \delta_{k} - \theta_{k})}$ and $\theta_{k}^{\prime} = {\mathcal H}\, \sin{(2 \delta_{k} - \theta_{k})}/(\cosh{r_{k}}\, \sinh{r_{k}})$. Their explicit solution is however not required to generalize the entropy bounds to the field theory set-up.}. It is however not essential, for the present purposes, to analyze these equations since the density matrices of the final state depend on the parametrization of Eq. (\ref{FT14}) but not on the specific functional form of the solution, exactly as in the 
quantum mechanical example discussed in the previous sections. 

\subsection{The total density operator of the final state}
For each $\vec{k}$-mode the basis for the irreducible representations of the 
$SU(1,1)$ group is analogous to the one previously introduced in the quantum mechanical analysis of the problem
with the difference that the dependence on the $\vec{k}$-mode must be consistently included:
\begin{equation}
| \,\{m^{({\mathcal s})}\}\,\, \{m^{({\mathcal e})}\,\} \rangle =\, \prod_{\vec{k}}\,\, |m_{\vec{k}}^{({\mathcal s})}\,\, m_{\vec{k}}^{({\mathcal e})} \rangle.
\label{SFT1}
\end{equation}
The states $|m_{\vec{k}}^{({\mathcal s})}\,\, m_{\vec{k}}^{({\mathcal e})} \rangle$ appearing in Eq. (\ref{SFT1})
are now  defined by:
\begin{equation}
|m_{\vec{k}}^{({\mathcal s})}\,\, m_{\vec{k}}^{({\mathcal e})} \rangle = \frac{(\widehat{{\mathcal s}}^{\,\,\dagger})^{m_{\vec{k}}^{({\mathcal s})}}}{\sqrt{m_{\vec{k}}^{({\mathcal s})}!}} \, \frac{(\widehat{{\mathcal e}}^{\,\,\dagger})^{m_{\vec{k}}^{({\mathcal e})}}}{\sqrt{m_{\vec{k}}^{({\mathcal e})}!}} |0_{\vec{k}}^{({\mathcal s})}\,\, 0_{\vec{k}}^{({\mathcal e})} \rangle.
\label{SFT2}
\end{equation}
The states $|m_{\vec{k}}^{({\mathcal s})}\,\, m_{\vec{k}}^{({\mathcal e})} \rangle$ form a convenient basis 
for the irreducible representations of $SU_{\vec{k}}(1,\,1)$ whose generators 
are now expressed as 
\begin{eqnarray}
&& \widehat{K}_{\vec{k}}^{(+)} = \widehat{{\mathcal s}}_{\vec{k}}^{\,\,\dagger}\, \widehat{{\mathcal e}}_{\vec{k}}^{\,\,\dagger}, \qquad 
\widehat{K}_{\vec{k}}^{(-)} = \widehat{{\mathcal s}}_{\vec{k}}\,\, \widehat{{\mathcal e}}_{\vec{k}},
\nonumber\\
&& \widehat{K}_{\vec{k}}^{(0)} = \bigl( \widehat{{\mathcal s}}_{\vec{k}}^{\,\,\dagger}\, \widehat{{\mathcal s}}_{\vec{k}} + \widehat{{\mathcal e}}_{\vec{k}}\,\,\widehat{{\mathcal e}}_{\vec{k}}^{\,\,\dagger}\bigr)/2.
\label{SFT3}
\end{eqnarray}
The commutation relations of the generators generalize the ones already discussed 
in Eqs. (\ref{OP1})--(\ref{OP2})
\begin{eqnarray}
&& [ \widehat{K}_{\vec{k}}^{(+)}, \widehat{K}_{\vec{p}}^{(-)}] = - 2 \widehat{K}_{\vec{k}}^{(0)}\, \delta_{\vec{k},\,\vec{p}}\,,
\nonumber\\ 
&& [ \widehat{K}_{\vec{k}}^{(0)}, \widehat{K}_{\vec{p}}^{(\pm)}] = \pm \widehat{K}_{\vec{k}}^{(\pm)}\, \delta_{\vec{k},\,\vec{p}}\,.
\label{SFT4}
\end{eqnarray}

In terms of the generators of Eq. (\ref{SFT3}) 
the operator $\widehat{\Sigma}_{\vec{k}}(z_{k})  = \exp{(z_{k}^{\ast} \, {\mathcal s}_{\vec{k}} {\mathcal e}_{\vec{k}} - z_{k} \, {\mathcal s}_{\vec{k}}^{\dagger} {\mathcal e}_{\vec{k}}^{\dagger} )}$ can be written in terms of the Baker-Campbell-Hausdorff decomposition
\begin{equation}
\widehat{\Sigma}_{\vec{k}}(z_{k}) = \widehat{\Sigma}^{(\vec{k})}_{+}(z_{k}) \, \widehat{\Sigma}^{(\vec{k})}_{0}(z_{k})
\, \widehat{\Sigma}^{(\vec{k})}_{-}(z_{k})
\label{SFT4a}
\end{equation}
 where $\widehat{\Sigma}^{(\vec{k})}_{\pm}(z_{\vec{k}})$ and $\widehat{\Sigma}^{(\vec{k})}_{0}(z_{\vec{k}})$ are 
 defined as:
\begin{eqnarray}
\widehat{\Sigma}^{(\vec{k})}_{-}(z_{\vec{k}}) &=& \exp{[ e^{- i \theta_{k}} \, \tanh{r_{r}}\, \widehat{K}_{\vec{k}}^{(-)}]} ,
\nonumber\\
&=& \exp{[(v_{k}^{\ast}/u_{k}^{\ast}) \widehat{K}_{\vec{k}}^{(-)}]}
\nonumber\\
\widehat{\Sigma}^{(\vec{k})}_{0}(z_{\vec{k}}) &=& \exp{[ - 2 \ln{ \cosh{r_{k}}} \,\widehat{K}_{\vec{k}}^{(0)}]},
\nonumber\\
&=&\exp{[ - 2 \ln{ |u_{k}|} \,\widehat{K}_{\vec{k}}^{(0)}]}
\nonumber\\
\widehat{\Sigma}^{(\vec{k})}_{+}(z_{\vec{k}}) &=& \exp{[ - e^{- i \theta_{k}} \, \tanh{r_{r}}\, \widehat{K}_{\vec{k}}^{(-)}]} 
\nonumber\\
&=&\exp{[- (v_{k}/u_{k}) \widehat{K}_{\vec{k}}^{(-)}]}.
\label{SFT4b}
\end{eqnarray}
In Eq. (\ref{SFT4b}) the Baker-Campbell-Hausdorff decomposition has been 
expressed both in terms of $(r_{k}, \,\theta_{k})$ and in terms of $(u_{k},\, v_{k})$
to clarify the absence of the $\delta_{k}$-dependence.

The analog of  $|\delta\, z\rangle$ introduced in Eq. (\ref{OP7}) is constructed 
from in analogy with the multiparticle Fock states of Eq. (\ref{SFT1}):
\begin{equation}
|\{\delta \,\,z\}\rangle = \prod_{\vec{k}} \,\,|\delta_{\vec{k}} \,\,z_{\vec{k}}\rangle,
\label{SFT5}
\end{equation}
where $|\delta_{\vec{k}} \,\,z_{\vec{k}}\rangle$ is now:
\begin{equation}
|\delta_{\vec{k}} \,\,z_{\vec{k}}\rangle= \widehat{{\mathcal R}}_{\vec{k}}(\delta_{\vec{k}})\,\,\widehat{\Sigma}_{\vec{k}}(z_{\vec{k}}) |0_{\vec{k}}^{({\mathcal s})}\,\, 0_{\vec{k}}^{({\mathcal e})} \rangle.
\end{equation}
Within the same formalism leading to Eqs. (\ref{SFT1})--(\ref{SFT2}) 
(\ref{SFT4})--(\ref{SFT5}) the initial state density matrix is
\begin{eqnarray}
\widehat{\rho}(\tau_{in}) &=& \prod_{\vec{k}} \widehat{\rho}_{\vec{k}}(\tau_{in})
\nonumber\\
&=& \prod_{\vec{k}} \widehat{\rho}_{\vec{k}}^{(s)}(\tau_{in}) \otimes 
\widehat{\rho}_{\vec{k}}^{(e)}(\tau_{in}).
\label{SFT6}
\end{eqnarray}
As in the quantum mechanical case (see Eq. (\ref{R1}) and discussions therein)
the initial density operator of the system is in the multiparticle vacuum (i.e.
$\widehat{\rho}_{\vec{k}}^{(s)}(\tau_{in}) =  |0_{\vec{k}}^{({\mathcal s})}\rangle \langle 0_{\vec{k}}^{({\mathcal s})}|$) whereas $\widehat{\rho}_{\vec{k}}^{(e)}(\tau_{in})$ is 
a thermal density matrix 
\begin{equation}
\widehat{\rho}_{\vec{k}}^{(e)}(\tau_{in})= \sum_{m_{\vec{k}}^{({\mathcal e})}=0}^{\infty} \, \overline{p}_{m_{\vec{k}}^{({\mathcal e})}} | m_{\vec{k}}^{({\mathcal e})}\rangle \langle m_{\vec{k}}^{({\mathcal e})}|,
\label{SFT7}
\end{equation}
characterized by a Bose-Einstein probability distribution for each of the 
$\vec{k}$-modes of the environment 
\begin{equation}
\overline{p}_{m_{\vec{k}}^{({\mathcal e})}} = \frac{\overline{n}_{k}^{\,\,m_{\vec{k}}^{({\mathcal e})}}}{(\overline{n}_{k}+1)^{m_{\vec{k}}^{({\mathcal e})}+1}},
\label{SFT8}
\end{equation}
where, as before, $\overline{n}_{k}$ indicates the averaged thermal multiplicity defined as 
\begin{equation}
\overline{n}_{k} = \frac{1}{e^{(\omega_{k} - \mu_{e})/T} -1}.
\label{SFT9}
\end{equation}
While in the quantum mechanical situation the average multiplicity did only depend 
on the single frequency $\omega_{e}$, Eq. (\ref{SFT9}) involves 
all the $\vec{k}$-modes associated with the environment. Putting together 
the results of Eqs. (\ref{SFT4})--(\ref{SFT5}) and of Eqs. (\ref{SFT6})--(\ref{SFT9})
the density matrix of the final state can then be obtained 
from the initial state as 
\begin{equation}
\widehat{\rho}(\tau_{fin},\tau_{in}) = \widehat{\mathcal U}(\{\delta\,\, z\} )\widehat{\rho}(\tau_{in}) \widehat{\Sigma}^{\dagger}(\{z\})\, \widehat{\mathcal U}^{\dagger}(\{\delta\,\, z\}).
\label{SFT10}
\end{equation}
where $ \widehat{\mathcal U}(\{\delta\,\, z\} )= \widehat{{\mathcal R}}(\{\delta\}) \widehat{\Sigma}(\{z\})$. The 
proper density operator encompassing the system and the environment can also be written as:
\begin{equation}
\widehat{\rho}(\tau_{fin},\tau_{in}) = \prod_{\vec{k}} \,\widehat{\rho}_{\vec{k}}(\tau_{fin},\tau_{in}).
\label{SFT10a}
\end{equation}
The explicit form of $\widehat{\rho}_{\vec{k}}(\tau_{fin},\tau_{in})$ becomes: 
\begin{eqnarray}
\widehat{\rho}_{\vec{k}}(\tau_{fin},\tau_{in}) &=&
\widehat{\mathcal U}_{\vec{k}}(\delta_{k},\,z_{k})\widehat{\rho}_{\vec{k}}(\tau_{in}) \, \widehat{\mathcal U}_{\vec{k}}^{\dagger}(\delta_{k}, z_{k})
\label{SFT10b}\\
 &=& \sum_{m_{\vec{k}}^{({\mathcal e})}=0}^{\infty} \sum_{\ell_{\vec{k}}=0}^{\infty} \sum_{\ell^{\prime}_{\vec{k}}=0}^{\infty} 
{\mathcal A}_{m_{\vec{k}}^{({\mathcal e})}\,\ell_{\vec{k}}\,\ell^{\prime}_{\vec{k}}} 
\nonumber\\
&\times& | \ell_{\vec{k}}, \ell_{\vec{k}} + m_{\vec{k}}^{({\mathcal e})} \rangle \langle m_{\vec{k}}^{({\mathcal e})} + \ell^{\prime}_{\vec{k}}, \ell^{\prime}_{\vec{k}}|.
\label{SFT11}
\end{eqnarray}
The term ${\mathcal A}_{m_{\vec{k}}^{({\mathcal e})}\,\ell_{\vec{k}}\,\ell^{\prime}_{\vec{k}}}$ has been included for convenience also with the purpose of exhibiting 
the analogy with the quantum mechanical situation of Eq. (\ref{R3}) 
\begin{equation}
{\mathcal A}_{m_{\vec{k}}^{({\mathcal e})}\,\ell_{\vec{k}}\,\ell^{\prime}_{\vec{k}}}= P_{m_{\vec{k}}^{({\mathcal e})}\ell_{\vec{k}} \ell^{\prime}_{\vec{k}}} \sqrt{\binom{m_{\vec{k}}^{({\mathcal e})}+ \ell_{\vec{k}}}{m_{\vec{k}}^{({\mathcal e})}} \binom{m_{\vec{k}}^{({\mathcal e})}+ \ell^{\prime}_{\vec{k}}}{m_{\vec{k}}^{({\mathcal e})}} },
\label{SFT12}
\end{equation}
where $P_{m_{\vec{k}}^{({\mathcal e})}\ell_{\vec{k}} \ell^{\prime}_{\vec{k}}}$ is
\begin{equation}
P_{m_{\vec{k}}^{({\mathcal e})}\ell_{\vec{k}} \ell^{\prime}_{\vec{k}}} =  \frac{e^{i \alpha_{\vec{k}} (\ell_{\vec{k}} -\ell^{\prime}_{\vec{k}})}\,\overline{p}_{m_{\vec{k}}^{({\mathcal e})}}}{(\cosh r_{k})^{2 (m_{\vec{k}}^{({\mathcal e})}+1) }} (\tanh{r_{k}})^{(\ell_{\vec{k}} +\ell^{\prime}_{\vec{k}})},
\label{SFT13}
\end{equation}
and $\alpha_{\vec{k}} = (\theta_{\vec{k}} +\pi -2 \delta_{\vec{k}})$.

\subsection{Extensions of the entropy bound}
From the total density matrix of Eqs. (\ref{SFT11}) and (\ref{SFT12})--(\ref{SFT13}) we now derive the reduced density matrices of the system and of the environment 
\begin{eqnarray}
&& \widehat{\rho}^{(e)}_{\vec{k}}(\tau, \tau_{in}) = \mathrm{Tr}_{s}[ \widehat{\rho}_{\vec{k}}(\tau,\, \tau_{in})],
\label{SFT14}\\
&& \widehat{\rho}^{(s)}_{\vec{k}}(\tau,\tau_{in}) = \mathrm{Tr}_{e}[ \widehat{\rho}_{\vec{k}}(\tau,\, \tau_{in})].
\label{SFT15}
\end{eqnarray}
The explicit form of $\widehat{\rho}^{(s)}_{\vec{k}}(\tau,\tau_{in})$ is:
\begin{equation}
\widehat{\rho}^{(s)}_{\vec{k}}(\tau,\tau_{in}) = \sum_{n_{\vec{k}}^{(e)}=0}^{\infty} 
\langle n_{\vec{k}}^{(e)}| \widehat{\rho}_{\vec{k}}(\tau,\, \tau_{in}) | n_{\vec{k}}^{(e)}\rangle.
\label{SFT16}
\end{equation}
Taking now into account the explicit results of Eqs. (\ref{SFT11}) and (\ref{SFT12})--(\ref{SFT13}) the reduced density matrix of the system becomes:
\begin{equation}
\widehat{\rho}^{(s)}_{\vec{k}}(\tau,\tau_{in}) = \sum_{\ell_{\vec{k}=0}}^{\infty} \frac{\overline{N}_{k}^{\ell_{\vec{k}}}}{(\overline{N}_{k} +1)^{\ell_{\vec{k}}+1}}\, 
|\ell_{\vec{k}} \rangle \langle \ell_{\vec{k}}|,
\label{SFT17}
\end{equation}
where $\overline{N}_{k}$ is now defined as:
\begin{equation}
\overline{N}_{k} = \overline{n}_{k}^{(q)} (\overline{n}_{k} +1) = (\overline{n}_{k} +1)\,\sinh^2{r_{k}}
\label{SFT18}
\end{equation}
The expression of $\overline{N}_{k}$ is fully analogous to the one 
already mentioned in Eq. (\ref{SS3}) with the difference that now all the multiplicities 
include an essential $k$-dependence. In particular Eq. (\ref{SFT18}) also implies that $\overline{N}_{k} \to 0$ as $r_{k} \to 0$: in this limit the system remains in the vacuum and the obtained entropy variation must vanish.

To verify this important point we go back to Eqs. (\ref{SFT15})--(\ref{SFT16}) and note that from the  reduced density matrix $\widehat{\rho}_{s}=\prod_{\vec{k}} \widehat{\rho}_{\vec{k}}^{(s)}$ the associated 
entropy follows from the von Neumann expression:
\begin{equation}
S_{s} = - \mathrm{Tr}\bigl[ \widehat{\rho}_{s} \, \ln{\widehat{\rho}_{s}}\bigr],
\label{ENTR1}
\end{equation}
where, according to the established notations, $S_{s}= S[ \widehat{\rho}_{s}]$.
But since $\widehat{\rho}_{s} = \prod_{\vec{k}} \widehat{\rho}_{\vec{k}}^{(s)}$
and $\mathrm{Tr}[ \widehat{\rho}_{\vec{k}}^{(s)}] =1$ the total entropy can be computed mode by mode, i.e.
\begin{equation}
S_{s} = \sum_{\vec{k}} \, S_{\vec{k}}^{(s)}, \quad S_{\vec{k}}^{(s)} = -  \mathrm{Tr}\bigl[ \widehat{\rho}_{\vec{k}}^{(s)} \, \ln{\widehat{\rho}_{\vec{k}}^{(s)}}\bigr].
\label{ENTR1a}
\end{equation}
From these expressions, recalling Eq. (\ref{SFT17}), 
we can immediately compute the total variation of the entropy of the system:
\begin{eqnarray}
\Delta S_{\vec{k}}^{(s)} &=& S_{\vec{k}}^{(s)}(\tau_{fin}) - S_{\vec{k}}^{(s)}(\tau_{in})
\nonumber\\
&=&  (\overline{N}_{k} + 1)\ln{(\overline{N}_{k} +1)} - \overline{N}_{k}\ln{\overline{N}_{k}}.
\label{ENTR2}
\end{eqnarray}
Recalling the comment mentioned after Eq. (\ref{SFT18}) we see from Eq. (\ref{ENTR2})
that $\Delta S_{\vec{k}}^{(s)}\to 0$ for $r_{k} \to 0$: this limit corresponds to the absence 
of parametric amplification (since $\overline{n}^{(q)}_{k} \to 0$) since under the condition 
$r_{k}\to 0$ the system remains, in practice, in the vacuum. 

As in the quantum mechanical case the total variation of the heat transferred to the environment can be computed either from the reduced density matrix $\widehat{\rho}_{e}$ (see Eq. (\ref{QQ2})  and discussion therein) or 
directly from the evolution of the multiplicities, as suggested in Eq. (\ref{QQ3}).
The same twofold possibility arises in the present context and the final result is:
\begin{eqnarray}
\Delta Q_{\vec{k}}^{(e)} &=& \langle \mathrm{th}| \widehat{H}_{\vec{k}}^{(e)}(\tau_{fin})|\mathrm{th} \rangle - \langle \mathrm{th}| \widehat{H}_{\vec{k}}^{(e)}(\tau_{in}) |\mathrm{th} \rangle 
\nonumber\\
&=& \omega_{k} \Delta N_{\vec{k}}^{(e)}, 
\label{ENTR3a}
\end{eqnarray}
where $|\mathrm{th}\rangle$ is the multiparticle state computed from the thermal density 
matrix where each mode of the field is characterized by a $k$-dependent Bose-Einstein 
probability distribution. Moreover in Eq. (\ref{ENTR3a})  $\Delta N_{\vec{k}}^{(e)}$ indicates the flow of particles 
to the environment and it is given by $\Delta N_{\vec{k}}^{(e)} = \overline{N}_{k} = \overline{n}_{k}^{(q)} (\overline{n}_{k} +1)$.

We can now verify, as anticipated, the generalized 
form of the bound already discussed at length in section \ref{sec5}: 
\begin{eqnarray}
T \Delta S_{\vec{k}}^{(s)} \leq \Delta Q_{\vec{k}}^{(e)} - \mu_{e} \Delta N_{\vec{k}}^{(e)}.
\label{ENTR5}
\end{eqnarray}
If we now recall the explicit expressions of Eqs. (\ref{ENTR2})--(\ref{ENTR3a}) for $\Delta S_{\vec{k}}^{(s)}$, $\Delta Q_{\vec{k}}^{(e)}$ and $\Delta N_{\vec{k}}^{(e)}$ we obtain that the  
bound of Eq. (\ref{ENTR5}) corresponds to 
\begin{equation}
\frac{( \overline{N}_{k} + 1)\ln{(\overline{N}_{k} +1)} - \overline{N}_{k}\ln{\overline{N}_{k}}}{\ln{(1 + 1/\overline{n}_{k})}} \leq 1.
\label{ENTR6}
\end{equation}
The condition (\ref{ENTR6}) generalizes the quantum mechanical result 
to the field theoretical situation and it can be analyzed in the same manner. 
The interesting aspect of Eq. (\ref{ENTR6}) is that $\overline{n}^{(q)}_{k}$ and $\overline{n}_{k}$ are now different 
for the various $k$-modes. This also shows that the entropy bounds may also depend on the underlying cosmological evolution \cite{NINE}. 

\subsection{The case of relic gravitons}
The evolution of the tensor modes of the geometry (corresponding to a massless field of spin $2$ evolving in curved space-times) shares various analogies with 
the class of problems analyzed in this investigation, as recently pointed out \cite{NINE}. 
The potential existence of stochastic backgrounds of relic gravitational radiation has been suggested even before the formulation of inflationary scenarios \cite{NINETEEN,TWENTY,TWENTYONE} as a genuine general relativistic effect in curved space-times. Since the evolution of the tensor modes of the geometry is not Weyl-invariant \cite{NINETEEN}, the associated classical and
quantum fluctuations can be amplified not only in anisotropic metric but also in conformally flat background geometries \cite{TWENTYONE} (see also \cite{TWENTYONEa}). For this reason backgrounds of relic gravitons
are expected, with rather different properties, in a variety of cosmological scenarios and, in particular, during an isotropic phase of quasi-de Sitter expansion \cite{TWENTYTWO,TWENTYTWOa,TWENTYTWOb}. The second-order tensor fluctuation of the Einstein-Hilbert action in a spatially flat Friedmann-Robertson-Walker background is given by \cite{TWENTYTWOc,TWENTYTWOd}
\begin{equation}
S_{g} = \frac{1}{8\ell_{P}^2} \int \, d^{4} x \,a^2(\tau) \, \eta^{\mu\nu}\, \partial_{\mu} \, h_{i\, j} \, \partial_{\nu} h^{i\, j},
\label{RG1}
\end{equation} 
where  $h_{i}^{\,\, i} = \partial_{i} h^{i\, j} =0$ describes the tensor modes of the geometry  while $\eta_{\mu\nu}$ denotes the Minkowski metric with signature $(+, -, -, -)$; as before $a(\tau)$ is the scale factor, written as a function of the conformal time coordinate $\tau$.  The rescaled canonical amplitudes and the comoving momenta analog to the 
ones introduced in the scalar case are now given by 
\begin{equation}
\mu_{i\, j}= h_{i\,j} a(\tau), \quad \pi_{i\,j} = (\partial_{\tau} \mu_{i\, j} - {\mathcal H} \mu_{i\, j})/(8\ell_{P}^2),
\label{RG2}
\end{equation}
where, as previously mentioned, ${\mathcal H} = a^{\prime}/a$ indicates the relative variation of the 
scale factor in terms of the conformal time coordinate.
During an inflationary stage of expansion the classical inhomogeneities 
are quickly ironed; the quantum fluctuations are described by 
the Hermitian operators $\widehat{\mu}_{i\, j}(\vec{x},\tau)$ and $\widehat{\pi}_{i\,j}(\vec{x},\tau)$
\begin{eqnarray}
&&\widehat{\mu}_{i\,j}(\vec{x},\tau) = \sqrt{2} \, \ell_{P}\, \int \frac{d^{3} k}{ (2 \pi)^{3/2}}\sum_{\alpha}\, e_{i\, j}^{(\alpha)} \widehat{\mu}_{\vec{k}, \, \alpha} e^{- i \vec{k}\cdot\vec{x}}, 
\nonumber\\
&&\widehat{\pi}_{i\,j}(\vec{x},\tau) = \frac{1}{4 \sqrt{2} \ell_{P}}\int  \frac{d^{3} k}{ (2 \pi)^{3/2}}\sum_{\alpha}\, e_{i\, j}^{(\alpha)}\widehat{\pi}_{\vec{k}, \, \alpha} e^{- i \vec{k}\cdot\vec{x}},
\nonumber
\end{eqnarray}
where the sums run over the two tensor polarizations $\alpha = \oplus, \otimes$, i.e.
\begin{equation}
e_{i\,j}^{(\oplus)}(\hat{k}) = (\hat{m}_{i} \, \hat{m}_{j} + \hat{n}_{i} \, \hat{n}_{j}),\quad
e^{(\otimes)}_{i\,j}(\hat{k}) = (\hat{m}_{i} \, \hat{n}_{j} - \hat{m}_{j} \, \hat{n}_{i}).
\label{RG2b}
\end{equation}
Note that, in Eq. (\ref{RG2b}), $\hat{m}$, $\hat{n}$ and $\hat{k}$ are just a triplet of mutually orthogonal unit vectors obeying $\hat{m} \times \hat{n}= \hat{k}$. In terms of the creation and annihilation operators  we have 
\begin{eqnarray}
\widehat{\mu}_{\vec{k}, \, \alpha} &=& ( \widehat{a}_{\vec{k},\, \alpha} + \widehat{a}_{-\vec{k},\, \alpha}^{\dagger})/\sqrt{2 \, k},
\nonumber\\
\widehat{\pi}_{\vec{k}, \, \alpha} &=& - i\, ( \widehat{a}_{\vec{k},\, \alpha} - \widehat{a}_{-\vec{k},\, \alpha}^{\dagger})\, \sqrt{k/2}.
\label{RG3}
\end{eqnarray}
Equation (\ref{RG3}) generalizes the result of Eq. (\ref{FT7}) to the case of the action (\ref{RG1}). Ultimately the Hamiltonian operator deduced from the action (\ref{RG1}) takes then the same form of Eq. (\ref{FT8})
\begin{eqnarray}
&& \widehat{H}_{g}(\tau) = \frac{1}{2} \int d^{3}k \sum_{\alpha} \, k\, \bigl[ \widehat{a}_{\vec{k},\,\alpha}^{\dagger} \widehat{a}_{\vec{k},\, \alpha} + 
\widehat{a}_{-\vec{k},\, \alpha}\widehat{a}_{-\vec{k},\, \alpha}^{\dagger}\bigr]
\nonumber\\
&&+ \frac{1}{2} \int d^{3}k \sum_{\alpha} \,\bigl[  \sigma\,
\widehat{a}_{\vec{k},\, \alpha}^{\dagger} \widehat{a}_{-\vec{k},\, \alpha}^{\dagger} + \sigma^{\ast} \,\widehat{a}_{\vec{k},\, \alpha} \widehat{a}_{-\vec{k},\, \alpha}  \bigr],
\label{RG4}
\end{eqnarray}
where $\sigma= i {\mathcal H}$. The previous quantum mechanical analysis is now 
easily extended to the case of the relic gravitons 
 by bearing in mind that the modes of the field with opposite three-momenta now operate in two different Hilbert subspaces \cite{NINE}. 
 
\renewcommand{\theequation}{7.\arabic{equation}}
\setcounter{equation}{0}
\section{Concluding remarks}
\label{sec7}
In a classical perspective the information erasure always requires an energy cost and, according 
to the current lore, this general conclusion should hold in spite of the actual size of the underlying physical structure.
If a certain system $s$ interacts with a thermal environment $e$
at a temperature $T$ the heat flowing to the environment $\Delta Q_{e}$
should always exceed $- T \Delta S_{s}$ where $\Delta S_{s}$ 
is the change of the entropy of the system between the final and the initial 
state. To erase one bit of information the entropy must then decrease as $\Delta S_{s} = - \ln{2}$ 
and the Landauer's conjecture would then imply that $\Delta Q_{e} \geq T \ln{2}$.
The Landauer's bound quantifies the 
minimum heat cost for obliterating information but the condition $\Delta Q_{e} \geq - T \, \Delta S_{s}$ is only restrictive if the entropy of the system decreases (i.e. $ \Delta S_{s} <0$) whereas in 
the opposite physical situation (i.e. $\Delta S_{s} \geq 0$) the increment of $S_{s}$ remains unconstrained. 
Since both conclusions should hold true for all physical structures (i.e. regardless of their respective sizes) 
we found appropriate to analyze them from a quantum mechanical viewpoint.

As the quantum regime is approached, the heat transfer and the entropy 
variations must be computed from a set of appropriately reduced density matrices.
While in statistical mechanics we typically deal with collections of many particles (of the 
order of the Avogadro number) in quantum thermodynamics we may even focus on a countable 
number of elementary oscillators. We then considered a minimal framework where both the system and the environment are described by two quantum oscillators with the difference that while the system is initially in a pure state, the environment is described by a mixed density operator with  Bose-Einstein (geometric) weights. We then  demonstrated that the lack of restrictions on the acquisition of the information 
disappears: if $\Delta S_{s} \geq 0$ the growth of the entropy of the system is bounded by the heat transferred to the environment according to the bound
\begin{equation}
\Delta S_{s} \leq \Delta Q_{e}/T,
\nonumber
\end{equation}
where $\Delta S_{s}$ is the entropy variation of the system, $\Delta Q_{e}$ is the 
heat flowing to the environment and $T$ is the corresponding temperature.
The Hermitian interaction between the system and the environment guarantees that the entropy of the system 
increases and if we put together the condition obtained here with the Landauer's bound we can then argue that 
\begin{equation}
- \Delta Q_{e}/T  \leq  \Delta S_{s} \leq \Delta Q_{e}/T.
\nonumber
\end{equation}
The inequality  $ \Delta S_{s} \leq \Delta Q_{e}/T$ 
expresses the bound discussed here while $- \Delta Q_{e}/T  \leq  \Delta S_{s}$ 
is just the standard form of the Landauer's conjecture of Eq. (\ref{INT1}). These two conditions could also be summarized by the requirement $T^2 (\Delta S_{s})^2 \leq (\Delta Q_{e})^2$.
The bound deduced here can also include  a particle flow to the environment (i.e. $\Delta N_{e} \geq 0$) associated with a chemical potential 
$\mu_{e} \neq 0$ 
\begin{equation}
 \Delta S_{s} \leq [\Delta Q_{e}/T - (\mu_{e}/T)\Delta N_{e}].
\nonumber
\end{equation}
When $\Delta S_{s} \geq 0$ the growth of the entropy of the system is bounded as $T \Delta S_{s} \leq (\Delta Q_{e} - \mu_{e}\Delta N_{e})$. There is finally a natural field-theoretical extension 
of the quantum mechanical considerations and it is related to the production of 
particles with spin $0$ or $2$. As we showed this generalization is also relevant for  
the relic gravitons in cosmological backgrounds.

Although from the classical viewpoint 
the acquisition of information and the related growth of the entropy of the system are never constrained, 
the quantum mechanical perspective pursued here demonstrates instead the opposite: the increase
of the entropy of the system is always limited by the heat transferred to the environment.
Since the basic features of our findings depend on the Hermitian nature of the interaction 
that should also amplify the entropy of the system, it would be interesting to characterize the general classes of physical structures for which an increase of the entropy of the system is always limited 
by the heat (and, possibly, by the particles) flowing to the environment.

\vspace{-0.5cm}
\section*{Acknowledgements}
\vspace{-0.5cm}
It is a pleasure to thank the kind assistance of A. Gentil-Beccot, A. Kohls, L. Pieper, S. Rohr and J. Vigen and of the whole CERN Scientific information Service.

\end{document}